\documentclass[11pt,a4paper]{article}
\usepackage[utf8]{inputenc}
\usepackage[T1]{fontenc}
\usepackage[english]{babel}
\usepackage{amsmath,amsfonts,amssymb}
\usepackage{graphicx}
\usepackage{float}
\usepackage{hyperref}
\usepackage[authoryear,round]{natbib}
\usepackage{authblk}
\usepackage[left=2.5cm,right=2.5cm,top=2.5cm,bottom=2.5cm]{geometry}
\usepackage{setspace}
\usepackage{abstract}
\usepackage{booktabs}
\usepackage{caption}
\usepackage{enumitem}
\usepackage{graphicx}

\title{\textbf{The dynamic of a tax on land value : concepts, models and impact scenario}}

\author[1]{Hugo Spring-Ragain\thanks{hugo.springragain@edu.ceds.fr // raguhugo37@gmail.com}}
\affil{CEDS, Paris}

\date{\today}

\begin{document}

\maketitle

\begin{abstract}
This paper develops a spatial–dynamic framework to analyze the theoretical and quantitative effects of a Land Value Tax (LVT) on urban land markets, capital accumulation, and spatial redistribution. Building upon the Georgist distinction between produced value and unearned rent, the model departs from the static equilibrium tradition by introducing an explicit diffusion process for land values and a local investment dynamic governed by profitability thresholds. Land value $V(x,y,t)$ and built capital $K(x,y,t)$ evolve over a two–dimensional urban domain according to coupled nonlinear partial differential equations, incorporating local productivity $A(x,y)$, centrality effects $\mu(x,y)$, depreciation $\delta$, and fiscal pressure $\tau$. 

Analytical characterization of the steady states reveals a transcritical bifurcation in the parameter $\tau$, separating inactive (low–investment) and active (self-sustaining) spatial regimes. The equilibrium pair $(V^*, K^*)$ is shown to exist only when the effective decay rate $\alpha = r + \tau - \mu(x,y)$ exceeds a profitability threshold $\theta = \kappa + \delta/I_0$, and becomes locally unstable beyond this boundary. The introduction of diffusion, $D_V \Delta V$, stabilizes spatial dynamics and generates continuous gradients of land value and capital intensity, mitigating speculative clustering while preserving productive incentives. Numerical simulations confirm these analytical properties and display the emergence of spatially heterogeneous steady states driven by urban centrality and local productivity.

The model also quantifies key aggregate outcomes, including dynamic tax revenues, adjusted capital-to-land ratios, and net present values under spatial heterogeneity and temporal discounting. Sensitivity analyses demonstrate that the main qualitative mechanisms—critical activation, spatial recomposition, and bifurcation structure—remain robust under alternative spatial profiles $(A, \mu)$, discretization schemes, and moderate differentiation of the tax rate $\tau(x,y)$. 

From an economic perspective, the results clarify the dual nature of the LVT: while it erodes unproductive rents and speculative land holding, its dynamic incidence on built capital depends on local profitability and financing constraints. The taxation parameter $\tau$ thus acts as a control variable in a nonlinear spatial system, shaping transitions between rent-driven and production-driven equilibria. Within a critical range around $\tau_c$, the LVT functions as an efficient spatial reallocation operator—reducing inequality in land values and investment density without impairing aggregate productivity. Beyond this range, excessive taxation induces systemic contraction and investment stagnation. 

Overall, this research bridges static urban tax theory with dynamic spatial economics by formalizing how a land-based fiscal instrument can reshape the geography of value creation through endogenous diffusion and nonlinear feedback. The framework provides a foundation for future extensions involving stochastic shocks, adaptive policy feedbacks, or endogenous public investment, offering a unified quantitative perspective on the dynamic efficiency and spatial equity of land value taxation.
\end{abstract}

\section*{Acknowledgements}
The author declares that this research received no specific grant from any funding agency in the public, commercial, or not-for-profit sectors.  
The entire manuscript was conceived, developed, and written solely by the author.  
The author also declares that there are no conflicts of interest related to the content of this article.

\newpage
\section{Introduction}

The question of land taxation occupies a central place in the economic analysis of urbanization, resource allocation, and fiscal equity. Land is a scarce, inelastic, and non-reproducible resource whose economic value derives primarily from location and collective amenities rather than from individual productive effort. In urban and peri-urban areas, this structural scarcity generates spatial concentration of value and the formation of substantial economic rents. Urbanization, the density of economic activity, the quality of public infrastructure, and the proximity to decision centers all contribute to the progressive capitalization of land value. This process is largely independent of the investments made by property owners, leading to a growing dissociation between value creation and productive effort.

In such a context, land speculation becomes a major source of distortion. The passive holding of land, in anticipation of future appreciation, allows owners to capture increasing rents driven by public externalities—such as infrastructure investment, transport network expansion, or local demographic dynamics. This phenomenon exerts persistent upward pressure on housing prices and contributes to a suboptimal allocation of land resources. By slowing the densification of central areas and deepening dependence on peripheral zones, it amplifies the social and environmental costs associated with urban sprawl while reducing the liquidity and efficiency of land and property markets.

Conventional tax systems are poorly equipped to address these dynamics. By focusing primarily on taxing income, consumption, or productive capital, they tend to neglect—or even indirectly favor—the passive capitalization of land value. This situation perpetuates a structural asymmetry between the sources of economic value and the tax base, fueling both allocative inefficiencies and perceptions of fiscal inequity. The absence of mechanisms directly targeting land rents encourages greater wealth concentration and speculative rather than productive land use.

It is precisely in this context that the Land Value Tax (LVT) holds a distinctive position in contemporary economic thought. Inspired by the seminal work of Henry George in \textit{Progress and Poverty}, the LVT proposes to tax the value of land independently of any building or improvement. By decoupling land from structures, the tax specifically targets unearned economic rent arising from location and collective externalities. It does not penalize productive investment but instead captures a portion of the value generated by factors exogenous to private ownership. The taxation of this rent rests on a fundamental economic principle: when a resource is perfectly inelastic, taxing it does not affect its supply or marginal allocation decisions, thereby granting the LVT a unique efficiency property within the field of public finance. It acts simultaneously as a mechanism to neutralize speculation—by reducing the incentive to hold idle land—and as a tool for redistributing the spatial rents derived from location advantages.

Despite the theoretical strength of these arguments, much of the literature on LVT remains confined to static frameworks. The dominant strand of research relies primarily on supply–demand models, institutional assessments, or short-term budgetary simulations. While these approaches capture the immediate effects of the tax, they often overlook essential dimensions linked to the temporal and spatial evolution of land values. They rarely incorporate, except marginally, the role of urban centrality, the spatial diffusion of values, the feedback between investment and profitability, or the delayed impact of taxation on the very structure of urban form.

These mechanisms are, however, inherently dynamic. The introduction of a land value tax alters the trajectory of land value accumulation by reshaping the expectations of landowners and investors. It acts on the intertemporal incentives to build, sell, or develop, and modifies the spatial patterns of attractiveness within the urban system. The effects of taxation cannot, therefore, be reduced to a static comparison between initial and final equilibria. They arise instead from a complex dynamic process in which land value diffuses through space according to neighborhood effects, gradients of centrality, and investment decisions that evolve over time. Understanding these mechanisms requires moving beyond the traditional static framework and adopting an explicit spatio-temporal perspective, capable of capturing the processes of diffusion, adaptation, and reallocation that shape the formation and evolution of land value within an urban economy.

\section{Theoretical fundamentals of the land value tax and litterature review}

The economic rationale of the Land Value Tax (LVT) rests on a fundamental conceptual distinction between produced value and economic rent. Extending the Georgist perspective, land rent is viewed as a collectively created form of wealth, dependent on location, public infrastructure, and agglomeration externalities rather than on the owner’s productive activity (George, 1879). In this framework, taxing land rent amounts to reallocating part of the value generated by the community back to the community, without reducing incentives to produce. This reasoning aligns with the classical principle of allocative neutrality: since the supply of land is inelastic, taxing it does not generate deadweight losses and thus constitutes an efficient source of public revenue.

These intuitions have been formalized in the optimal taxation literature. Foundational works by Vickrey (1977), Mirrlees (1971), and Atkinson and Stiglitz (1976) demonstrated that the taxation of pure rents—such as land rent—can yield substantial revenues without distorting economic efficiency. More recent contributions have nuanced this conclusion, emphasizing that the optimal level of rent taxation depends on redistributive objectives and on the way fiscal revenues are recycled (Best and Kleven, 2022; Kopczuk, 2022). Nonetheless, these studies converge on the view that the land tax base is among the most efficient available, which gives the LVT a unique status in the fiscal landscape.

Contemporary research has also focused on the ability of the LVT to capture land value increments generated by public investment, through mechanisms known as \emph{land value capture}. This principle relies on the idea that public spending on transport, infrastructure, or urban amenities becomes capitalized into land prices, and that taxing this added value allows for endogenous financing of public investment (Peterson, 2009; Grimes and Liang, 2021; Cheshire and Hilber, 2022). Recent empirical analyses show that such mechanisms produce stable and predictable fiscal revenues, even under macroeconomic volatility (Ingram and Hong, 2023; IMF, 2024). These properties strengthen the appeal of the LVT as both a sustainable public finance instrument and a local budget stabilizer.

From the perspective of urban economics, the integration of the LVT fits within the analytical tradition established by Alonso (1964), Mills (1967), and Muth (1969), who highlighted the structuring role of land rent in shaping urban morphology. In these models, land rent and the center–periphery gradient determine density, spatial allocation, and relative land prices. The introduction of a land tax alters these equilibria by reducing incentives for speculative landholding and redirecting investment toward high-productivity urban areas (Oates and Schwab, 1997; Dye and England, 2010). More recent studies confirm that increasing the share of the land tax relative to the building tax encourages construction and densification while flattening land price gradients (Banzhaf and Lavery, 2022; Gyourko and Sieg, 2023).

Empirical experiences across jurisdictions provide further support for these theoretical results. The differentiated tax systems between land and buildings implemented in Pennsylvania, for example, have shown that raising the land share in the tax base increases construction intensity and urban density (Plassmann and Tideman, 2022). Similarly, studies of the Estonian land reform and the Scandinavian fiscal frameworks highlight the robustness of land-based revenues and their ability to finance urban infrastructure while limiting speculative rent accumulation (Estonian Ministry of Finance, 2023; IMF, 2024).

Parallel to these developments, a more recent strand of literature examines the macroeconomic and distributive effects of land rent and its taxation. Niepelt (2014) and Rognlie (2015) showed that land rent accounts for a growing share of contemporary wealth inequality, and that land taxation can operate as an efficient redistributive tool without introducing distortions. These conclusions are reinforced by empirical studies documenting the temporal stability of land tax revenues and their contribution to local public finance even under uncertain macroeconomic conditions (OECD, 2023; IMF, 2024).

Despite the abundance of research, a common limitation in nearly all studies lies in their static framework. Theoretical analyses typically rely on comparative-static equilibrium models, while empirical studies assess the effects of LVT introduction or reform over relatively short horizons. Yet, land value formation is a dynamic process, characterized by spatial diffusion, gradual capitalization of infrastructure, and temporal feedbacks between fiscal incentives and investment decisions. This dynamic nature is at the core of several recent contributions emphasizing the need to incorporate an explicit temporal dimension to fully understand the effects of the LVT (Banzhaf et al., 2023; Glaeser and Gyourko, 2023; OECD, 2024).

\section{Mathematical models and results}
\subsection{Land value and built capital}

We consider an urban agglomeration represented by a rectangular spatial domain
\begin{equation}
\Omega = [0, L_x] \times [0, L_y], 
\end{equation}

where $L_x$ and $L_y$ denote the spatial dimensions along the $x$ and $y$ directions, respectively. The urban center is assumed to be located at coordinates $(L_x/2,\, L_y/2)$.

Within this domain, we define two fundamental variables that depend on both space $(x,y)$ and time $t$:
\begin{itemize}
    \item $V(x,y,t)$ : the \textit{land value}, excluding construction costs;
    \item $K(x,y,t)$ : the \textit{built capital stock}.
\end{itemize}

The core parameters and functional components of the model are as follows:
\[
\begin{aligned}
&r > 0 &&\text{: discount rate},\\
&\tau \geq 0 &&\text{: ad valorem tax rate applied to land value},\\
&D_V > 0 &&\text{: diffusion coefficient of land value},\\
&\mu(x,y) &&\text{: local growth effect function modifying the decay of value},\\
&A(x,y) > 0 &&\text{: local productivity determining the capacity to generate land rent},\\
&\beta \in (0,1) &&\text{: elasticity parameter measuring the contribution of built capital to production},\\
&c_b > 0 &&\text{: baseline construction cost},\\
&I_0 > 0 &&\text{: sensitivity coefficient of investment to profitability},\\
&\kappa > 0 &&\text{: minimum profitability threshold},\\
&\delta > 0 &&\text{: depreciation rate of built capital}.
\end{aligned}
\]

To incorporate the spatial effect of centrality, we let both $A(x,y)$ and $\mu(x,y)$ vary with the distance to the urban center. Defining the radial distance from the center as
\begin{equation}
r_{\text{cent}}(x,y) = \sqrt{(x - L_x/2)^2 + (y - L_y/2)^2},  
\end{equation}

we specify the following exponential profiles:

\begin{align}
A(x,y) &= A_0 \, \exp\big(-\gamma\, r_{\text{cent}}(x,y)\big),\\[3pt]
\mu(x,y) &= \mu_0 \, \exp\big(-\lambda\, r_{\text{cent}}(x,y)\big),
\end{align}

where $A_0$, $\mu_0$, $\gamma$, and $\lambda$ are positive constants.

These spatially decaying functions capture the intuition that both productivity and local growth potential are highest near the city center and gradually decrease toward the periphery. This spatial heterogeneity forms the backbone of the model’s dynamics, linking urban form to the diffusion of land value and investment incentives across space.

\subsubsection{Land value formation}

The land value function $V(x,y,t)$ arises from the combination of three main effects: discounting and taxation, spatial diffusion, and the productive contribution of built capital. These mechanisms jointly determine the spatio-temporal evolution of urban land value.\\

\textit{Discounting and taxation effects.} In the absence of incoming flows, the land value naturally decays at an effective rate $r$, reflecting intertemporal discounting. The imposition of a Land Value Tax (LVT) at rate $\tau$ further increases this rate of decay. However, a local growth effect, denoted $\mu(x,y)$, partly offsets this decline by capturing endogenous drivers of urban expansion or agglomeration effects. The combined term describing this net decay process is thus given by:
\begin{equation}
-[r + \tau - \mu(x,y)]\, V(x,y,t).    
\end{equation}\\

\textit{Spatial diffusion.} Spatial interactions between neighboring areas are modeled through a diffusion process represented by the Laplacian operator. The diffusion term $D_V \Delta V(x,y,t)$ captures the spatial smoothing of land values across locations, reflecting spillover and neighborhood effects. The two-dimensional Laplacian is defined as:
\begin{equation}
\Delta V(x,y,t) = \frac{\partial^2 V}{\partial x^2}(x,y,t) + \frac{\partial^2 V}{\partial y^2}(x,y,t),
\end{equation}

where $D_V > 0$ denotes the diffusion coefficient, measuring the intensity of spatial interdependence between locations.\\

\textit{Rent production by built capital.} Built capital $K(x,y,t)$ generates local rent, representing the productive utilization of land through development and construction. This contribution is modeled as:
\begin{equation}
A(x,y)\, K(x,y,t)^{\beta},
\end{equation}

where $A(x,y)$ is the local productivity term, and $\beta \in (0,1)$ represents the elasticity of land rent with respect to built capital.\\

\textit{Dynamic equation for land value.} Combining the above mechanisms, the spatio-temporal evolution of land value is governed by the following partial differential equation:
\begin{equation}
\frac{\partial V(x,y,t)}{\partial t}
= -[r + \tau - \mu(x,y)]\, V(x,y,t)
+ D_V \Delta V(x,y,t)
+ A(x,y)\, K(x,y,t)^{\beta}. 
\end{equation}

In this equation:
\begin{itemize}
    \item The term $-[r + \tau - \mu(x,y)] V(x,y,t)$ captures the joint effects of discounting and taxation, partially offset by local growth dynamics;
    \item The diffusion term $D_V \Delta V(x,y,t)$ redistributes land value spatially, accounting for neighborhood interactions and the smoothing of local shocks;
    \item The production term $A(x,y) K(x,y,t)^{\beta}$ injects a positive flow of value, proportional to the productive use of built capital.
\end{itemize}

Together, these components define a spatially extended, dynamic valuation process in which the evolution of land value depends simultaneously on temporal decay, spatial diffusion, and endogenous accumulation mechanisms.

\subsubsection{Built capital accumulation}

The built capital stock $K(x,y,t)$ evolves through the accumulation of net investment, which depends on local profitability conditions. The instantaneous profitability ratio is defined as:
\begin{equation}
\Pi(x,y,t) = \frac{A(x,y)\, K(x,y,t)^{\beta}}{V(x,y,t) + c_b}, 
\end{equation}

where the denominator $V(x,y,t) + c_b$ represents the total land value augmented by the baseline construction cost $c_b$. This formulation captures the idea that profitable investment occurs only when the expected return on construction exceeds the combined value of land acquisition and building costs.

\textit{Investment dynamics and depreciation.}  
The central assumption of the model is that the rate of investment is proportional to the existing stock of capital $K(x,y,t)$ and becomes effective only when local profitability exceeds a minimum threshold $\kappa > 0$. Moreover, a proportional depreciation process continuously erodes the capital stock at rate $\delta > 0$. Combining these mechanisms yields the following dynamic equation:
\begin{equation}
\frac{\partial K(x,y,t)}{\partial t}
= I_0 \left(\frac{A(x,y)\, K(x,y,t)^{\beta}}{V(x,y,t) + c_b} - \kappa\right) K(x,y,t)
- \delta\, K(x,y,t),   
\end{equation}

where $I_0 > 0$ denotes the sensitivity of investment to local profitability.

\textit{Interpretation.}  
In this formulation, the term
\begin{equation}
I_0 \left(\frac{A(x,y)\, K(x,y,t)^{\beta}}{V(x,y,t) + c_b} - \kappa\right) K(x,y,t) 
\end{equation}

represents the flow of net investment. It is positive when local profitability $\Pi(x,y,t)$ exceeds the threshold $\kappa$, driving accumulation of built capital, and negative when profitability falls below this threshold, leading to disinvestment or stagnation. The term $\delta\, K(x,y,t)$, by contrast, captures the continuous depreciation of existing structures due to aging, obsolescence, or wear.

\textit{Economic significance.}  
This equation formalizes a dynamic feedback loop between land value and capital formation. A high $A(x,y)$ or low $V(x,y,t)$ increases profitability, stimulating investment and urban densification. Conversely, high land values or low productivity reduce incentives to build, generating underutilized or vacant areas. The interaction between $V(x,y,t)$ and $K(x,y,t)$ thus drives the endogenous spatial evolution of urban form, making the local equilibrium between rent extraction and capital accumulation central to the overall dynamics of the system.

\subsubsection{Conditions and simulation}

To ensure a well-posed problem, the model requires both initial and boundary conditions for the coupled variables $V(x,y,t)$ and $K(x,y,t)$.

\textit{Initial conditions.}  
At the initial time $t = 0$, we specify the spatial distribution of both land value and built capital as:
\begin{equation}
\begin{cases}
V(x,y,0) = V_0(x,y), \\
K(x,y,0) = K_0(x,y),
\end{cases}    
\end{equation}

where $V_0(x,y)$ and $K_0(x,y)$ denote the initial states of land value and capital across the urban domain. These functions may represent, for example, an empirical calibration based on observed spatial heterogeneity or a theoretical steady-state profile prior to taxation.

\textit{Boundary conditions.}  
At the boundaries of the domain $\partial \Omega$, we impose homogeneous Neumann conditions to prevent artificial inflows or outflows of value and capital. This ensures that the spatial diffusion process remains confined within the urban system:
\begin{equation}
\begin{cases}
\displaystyle \frac{\partial V}{\partial n}\Big|_{\partial \Omega} = 0, \\
\displaystyle \frac{\partial K}{\partial n}\Big|_{\partial \Omega} = 0,
\end{cases}  
\end{equation}

where $\partial / \partial n$ denotes the outward normal derivative on the boundary. These reflective boundary conditions maintain spatial consistency and ensure mass conservation for both $V$ and $K$.

\textit{Numerical approximation.}  
For numerical exploration of the system, we adopt a two-dimensional finite-difference discretization scheme. The spatial derivatives are approximated using uniform steps $\Delta x$ and $\Delta y$, while time integration is performed through an explicit Euler method with a time step $\Delta t$. This numerical approach provides a straightforward and transparent framework for simulating the joint evolution of land value and built capital under various taxation scenarios, enabling a direct analysis of spatial propagation, equilibrium convergence, and transient dynamics.

\subsection{Value and rations of the land value and the built capital}

\subsubsection{Tax revenues}

The annual fiscal revenue generated by the introduction of a Land Value Tax (LVT) can be expressed in its simplest form as:
\begin{equation}
R_{\text{tax}}(t) = t \int_{\Omega} V(x,y,t)\, dx\, dy,    
\end{equation}

where $t$ denotes the ad valorem tax rate and $V(x,y,t)$ represents the spatially distributed land value at time $t$.  
This formulation, however, remains essentially static—it measures the instantaneous flow of tax revenue without accounting for the temporal evolution of land values induced by the LVT itself.

\textit{Dynamic formulation.}  
To capture the temporal dimension of fiscal returns, we introduce a dynamic measure of revenue that integrates the future trajectories of land value over a finite planning horizon $T$. The government collects tax revenues continuously while discounting future inflows at the rate $r$. Denoting by $V(x,y,t+\tau)$ the land value at future time $t+\tau$ for $\tau \in [0,T]$, the discounted present value of total fiscal revenue becomes:
\begin{equation}
R_{\text{tax}}^{AD}(t)
= \int_0^T e^{-r\tau}
\left[
t \int_{\Omega} V(x,y,t+\tau)\, w_1(x,y)\, dx\, dy
\right] d\tau,    
\end{equation}

where $w_1(x,y)$ is a spatial weighting function that can represent, for instance, administrative efficiency, differential collection rates, or regional fiscal priorities.\\

This expression captures the cumulative discounted revenue stream over time, accounting for both the endogenous dynamics of land value and the spatial heterogeneity of fiscal incidence.  
The introduction of $w_1(x,y)$ allows for flexible modeling of policy design — for example, weighting revenue more heavily in high-density or strategic zones. The integral formulation provides a comprehensive measure of the fiscal return from the LVT, consistent with dynamic general equilibrium approaches in spatial public finance. It also establishes a direct bridge between the temporal path of land capitalization and long-term public resource availability.

\subsubsection{Mean ponderate value for the land}

The average land value can be refined to reflect spatial heterogeneity by introducing a local weighting scheme.  
Rather than computing a simple spatial mean, we consider a weighted formulation that emphasizes economically or demographically active zones. Let $p(x,y)$ denote a spatial density function—such as population, built density, or transaction intensity—and $w_2(x,y)$ an additional spatial weight capturing policy relevance or data reliability. The weighted mean land value at time $t$ is then defined as:
\begin{equation}
\overline{V}(t)
= \frac{
\displaystyle \int_{\Omega} p(x,y)\, V(x,y,t)\, w_2(x,y)\, dx\, dy
}{
\displaystyle \int_{\Omega} p(x,y)\, w_2(x,y)\, dx\, dy }.   
\end{equation}\\

This formulation provides a more representative indicator of the average land value by emphasizing regions that contribute most to economic activity or fiscal interest.  
Unlike an unweighted spatial mean, $\overline{V}(t)$ adapts to the urban structure by assigning greater influence to areas with higher land-use intensity or policy significance.  
In dynamic simulations, $\overline{V}(t)$ serves as a key observable for tracking localized responses to taxation, investment, and centrality gradients, capturing short-term spatial variations in value without being dominated by peripheral or low-activity regions.

\subsubsection{Adjusted capital / land ratio}

A key indicator for assessing the interaction between built capital and land valuation is the ratio of total built capital to total land value.  
In its simplest form, this ratio is expressed as:
\begin{equation}
R_{K/V}(t)
= \frac{
\displaystyle \int_{\Omega} K(x,y,t)\, dx\, dy
}{
\displaystyle \int_{\Omega} V(x,y,t)\, dx\, dy }.    
\end{equation}

This basic formulation provides a global measure of how much productive capital has been accumulated relative to the aggregate value of land.  
However, it does not account for spatial heterogeneity or for differences in investment intensity across urban areas.

\textit{Adjusted formulation.}  
To capture these effects more precisely, we define an adjusted ratio that introduces both spatial and temporal weights, as well as a correction based on an investment intensity index $I(x,y)$:
\begin{equation}
R_{K/V}^{\text{adj}}(t)
= \frac{
\displaystyle \int_{\Omega} I(x,y)\, K(x,y,t)\, w_3(x,y)\, dx\, dy
}{
\displaystyle \int_{\Omega} I(x,y)\, V(x,y,t)\, w_3(x,y)\, dx\, dy }.    
\end{equation}

Here, $w_3(x,y)$ is a spatial weighting function that may represent, for instance, infrastructure accessibility or fiscal importance, while $I(x,y)$ quantifies the relative strength of local investment activity. \\

The adjusted ratio $R_{K/V}^{\text{adj}}(t)$ provides a more accurate picture of the relative intensity of built investment with respect to land valorization in active economic zones.  
By incorporating $I(x,y)$ and $w_3(x,y)$, the indicator highlights regions where the LVT most effectively stimulates productive investment rather than speculative retention.  
It thus serves as a dynamic diagnostic tool for evaluating how taxation policies reshape the balance between construction, land use, and spatial value accumulation over time.

\subsubsection{Adjusted land profitability and spatial aggregation}

The baseline formulation of local land profitability is defined as the ratio between the productive output of built capital and the land value on a given site:
\begin{equation}
Y(x,y,t) = \frac{A(x,y)\, K(x,y,t)^{\beta}}{V(x,y,t)}.    
\end{equation}

This expression captures the instantaneous efficiency of capital utilization relative to land value. However, it assumes perfect homogeneity in local conditions, ignoring both risk differentials and qualitative heterogeneity across space.

\textit{Adjusted formulation.}  
To provide a more realistic measure, we introduce a local risk factor $\sigma(x,y)$ and a multiplicative adjustment through a spatial quality function $q(x,y)$. The adjusted profitability becomes:
\begin{equation}
Y^{\text{adj}}(x,y,t)
= \frac{A(x,y)\, K(x,y,t)^{\beta}}{V(x,y,t)}
\cdot
\left(1 - \frac{\sigma(x,y)}{1 + q(x,y)}\right).    
\end{equation}

The term $\left(1 - \frac{\sigma(x,y)}{1 + q(x,y)}\right)$ moderates profitability according to risk exposure and local quality—reducing returns in high-risk areas while amplifying them in zones characterized by superior amenities or public infrastructure.

\textit{Spatially aggregated indicator.}  
The mean adjusted profitability over the entire domain $\Omega$ is computed through a weighted integration:
\begin{equation}
\overline{Y}^{\text{adj}}(t)
= \frac{
\displaystyle \int_{\Omega} w(x,y)\, Y^{\text{adj}}(x,y,t)\, dx\, dy
}{
\displaystyle \int_{\Omega} w(x,y)\, dx\, dy }.    
\end{equation}

Here, $w(x,y)$ denotes a spatial weighting function that can incorporate factors such as population density, land-use intensity, or policy relevance.\\

The indicator $\overline{Y}^{\text{adj}}(t)$ provides a synthetic measure of average land profitability, corrected for local quality and risk heterogeneity.  
It allows for an explicit assessment of how taxation and investment policies affect not only the aggregate efficiency of capital use but also its spatial distribution across heterogeneous urban environments.  
This formulation thus bridges local microeconomic profitability and macro-spatial performance within the LVT dynamic framework.

\subsubsection{Dynamic Net Present Value (NPV)}

\textit{Conceptual foundation.}  
To obtain a robust measure of the financial attractiveness of land investment under taxation, we define the \textit{dynamic net present value} (NPV) by integrating discounted net cash flows over a finite horizon $T$.  
This approach captures both the temporal and spatial heterogeneity of returns while accounting for local variations in discount rates and taxation levels.

\textit{Local formulation.}  
Let the local discount rate be given by
\begin{equation}
r(x,y) = r + \rho(x,y),    
\end{equation}

where $\rho(x,y)$ denotes a regional risk premium capturing spatial differences in uncertainty or credit conditions.  
The local net present value is then defined as:
\begin{equation}
\text{NPV}(x,y,t)
= \int_{0}^{T} e^{-r(x,y)\tau}
\Big[ R(x,y,t+\tau) - t\,V(x,y,t+\tau) \Big]\, d\tau, 
\end{equation}

where the term
\begin{equation}
R(x,y,t+\tau) = A(x,y)\, K(x,y,t+\tau)^{\beta}    
\end{equation}

represents the local rental or return flow generated by the built capital.  
The subtraction of $tV(x,y,t+\tau)$ accounts for the fiscal outflow induced by the land value tax, ensuring that the NPV measures the net financial benefit after taxation.

\textit{Spatial aggregation.}  
The mean NPV over the entire domain $\Omega$ is obtained by a weighted spatial integration:
\begin{equation}
\overline{\text{NPV}}(t)
= \frac{
\displaystyle \int_{\Omega} \text{NPV}(x,y,t)\, w_4(x,y)\, dx\, dy
}{
\displaystyle \int_{\Omega} w_4(x,y)\, dx\, dy }.    
\end{equation}

Here, $w_4(x,y)$ is a spatial weighting function that can reflect the distribution of population, fiscal significance, or development potential.\\

This dynamic formulation allows for a continuous evaluation of how land taxation affects the temporal profile of investment incentives.  
Regions characterized by high $\rho(x,y)$ experience stronger discounting, reflecting greater perceived risk or lower access to credit, while zones with high productivity $A(x,y)$ or dense capital $K(x,y,t)$ maintain positive NPV trajectories even under stronger taxation.  
The aggregate indicator $\overline{\text{NPV}}(t)$ thus provides an integrated measure of the system’s financial viability, linking spatial heterogeneity, risk-adjusted profitability, and fiscal sustainability within the Land Value Tax framework.

\subsection{Diffusion and Temporal Dimension}

We now consider a two-dimensional spatial domain:
\begin{equation}
\Omega = [0, L_x] \times [0, L_y],    
\end{equation}

where $L_x$ and $L_y$ denote the spatial dimensions along the $x$ and $y$ directions, respectively.  
The spatial grid is divided into $N_x$ points along $x$ and $N_y$ points along $y$, with spatial steps:
\begin{equation}
dx = \frac{L_x}{N_x - 1}, \quad dy = \frac{L_y}{N_y - 1}.    
\end{equation}

The grid points are defined as:
\begin{equation}
x_i = i\, dx, \quad i = 0, 1, \ldots, N_x - 1, \qquad
y_j = j\, dy, \quad j = 0, 1, \ldots, N_y - 1.    
\end{equation}

The continuous functions $V(x,y,t)$ and $K(x,y,t)$ are approximated by discrete matrices
$V_{i,j}(t)$ and $K_{i,j}(t)$ evaluated at the grid points $(x_i, y_j)$.

\textit{Spatial diffusion term.}  
The diffusive term in the land value equation involves the Laplacian $\Delta V$ defined by:
\begin{equation}
\Delta V(x,y,t) = \frac{\partial^2 V}{\partial x^2}(x,y,t) + \frac{\partial^2 V}{\partial y^2}(x,y,t), 
\end{equation}

which is discretized using central differences at point $(i,j)$ as:
\begin{equation}
\frac{\partial^2 V}{\partial x^2}\Big|_{i,j} \approx
\frac{V_{i+1,j} - 2V_{i,j} + V_{i-1,j}}{dx^2},
\qquad
\frac{\partial^2 V}{\partial y^2}\Big|_{i,j} \approx
\frac{V_{i,j+1} - 2V_{i,j} + V_{i,j-1}}{dy^2}.    
\end{equation}

Hence, the full discrete Laplacian is given by:
\begin{equation}
\Delta V_{i,j} \approx
\frac{V_{i+1,j} - 2V_{i,j} + V_{i-1,j}}{dx^2}
+ \frac{V_{i,j+1} - 2V_{i,j} + V_{i,j-1}}{dy^2}.    
\end{equation}

\textit{Temporal discretization.}  
Time is discretized using an explicit Euler scheme with time step $dt$.  
The temporal interval $[0, T_{\text{final}}]$ is divided into $N_t$ steps:
\begin{equation}
N_t = \frac{T_{\text{final}}}{dt},    
\end{equation}

and each time step is denoted $t^n = n\, dt$ for $n = 0, 1, \ldots, N_t$.  
At each step, $V(x,y,t)$ and $K(x,y,t)$ are approximated by $V_{i,j}^n$ and $K_{i,j}^n$.

\textit{Discrete evolution equations.}  
The land value equation becomes:
\begin{equation}
V_{i,j}^{n+1} =
V_{i,j}^n + dt \left[
-(r + \tau - \mu_{i,j})\, V_{i,j}^n
+ D_V\, \Delta V_{i,j}^n
+ A_{i,j}\, (K_{i,j}^n)^{\beta}
\right],    
\end{equation}

with
\begin{equation}
\mu_{i,j} = \mu(x_i, y_j), \quad A_{i,j} = A(x_i, y_j).    
\end{equation}

The built capital evolves according to:
\begin{equation}
K_{i,j}^{n+1} =
K_{i,j}^n + dt \left[
I_0
\left(
\frac{A_{i,j}\, (K_{i,j}^n)^{\beta}}{V_{i,j}^n + c_b}
- \kappa
\right)
K_{i,j}^n
- \delta\, K_{i,j}^n
\right].    
\end{equation}

\begin{figure}[H]
    \centering
    \includegraphics[width=\textwidth]{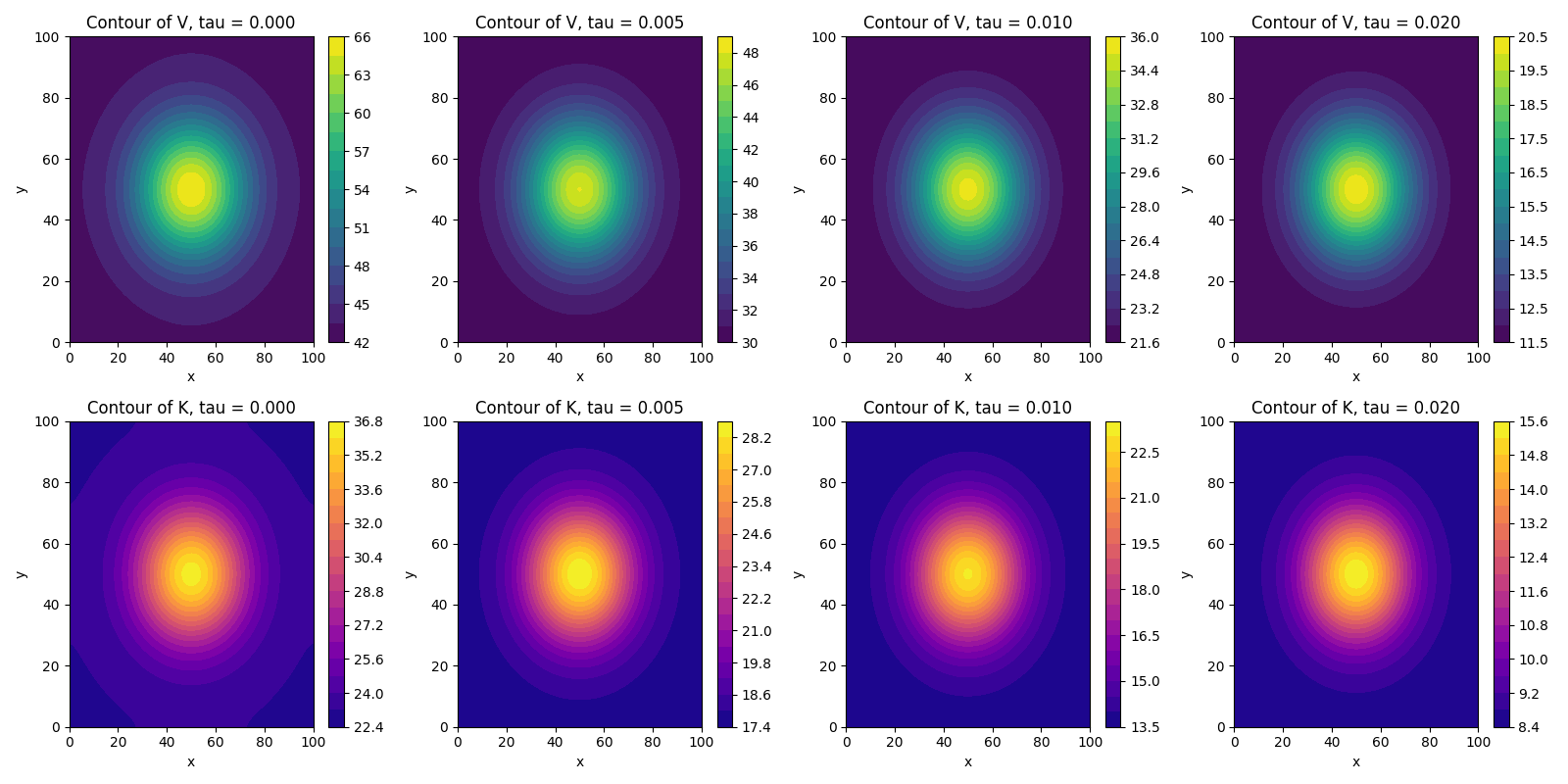}
    \caption{
        \textbf{Spatial diffusion and dynamic equilibrium under varying tax rates.}  
        The figure shows the contour maps of land value $V(x,y)$ (top row) and built capital $K(x,y)$ (bottom row) at the final steady state of the simulation for four tax levels $\tau \in \{0.0,\, 0.005,\, 0.01,\, 0.02\}$.  
        For $\tau = 0$, both $V$ and $K$ concentrate strongly around the domain center, reproducing the initial Gaussian peak.  
        As $\tau$ increases, the peak amplitude decreases significantly, indicating a progressive attenuation of land and capital accumulation.  
        Higher tax rates reduce profitability and thus slow the concentration of built capital, leading to a more homogeneous spatial distribution, especially evident in $K(x,y)$.  
        This pattern illustrates the erosion of built capital intensity under stronger land taxation within the diffusion–interaction framework.
    }
\end{figure}
\textit{Temporal evolution of spatial averages.}  
The first figure below presents the time evolution of the spatial averages of land value $V(x,y,t)$ and built capital $K(x,y,t)$ for several land value tax (LVT) rates $\tau$.  
Each pair of curves corresponds to the same tax rate, with solid lines representing $\langle V \rangle$ and dashed lines representing $\langle K \rangle$.  

\begin{figure}[H]
    \centering
    \includegraphics[width=0.8\textwidth]{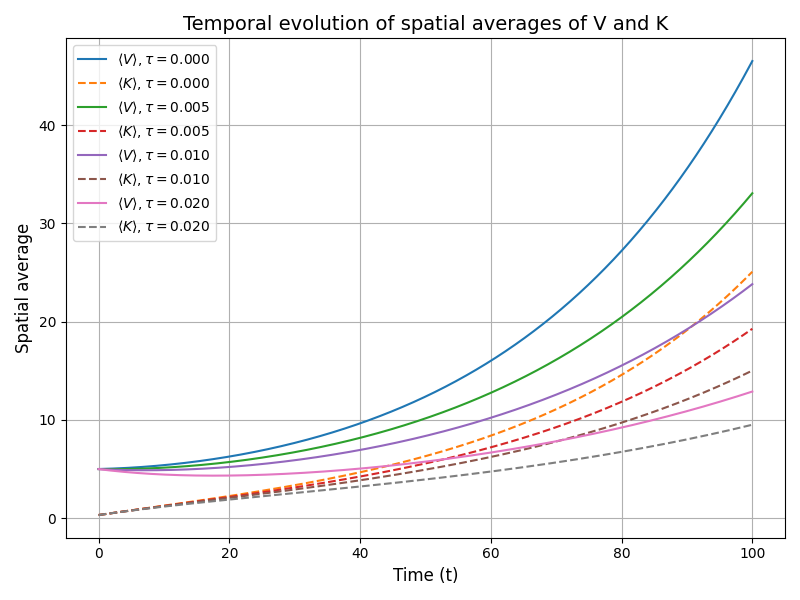}
    \caption{
        \textbf{Temporal evolution of spatial averages of $V$ and $K$ under different LVT rates.}  
        Each pair of curves corresponds to the same tax level $\tau$, with solid lines representing the average land value $\langle V \rangle$ and dashed lines the average built capital $\langle K \rangle$.  
        When $\tau = 0$, both $V$ and $K$ exhibit strong and sustained growth over time, reflecting an unconstrained accumulation dynamic in the absence of taxation.  
        As $\tau$ increases, the growth trajectories flatten markedly, revealing the damping effect of the land value tax on both capital accumulation and land appreciation.  
        Higher tax rates progressively reduce the slopes of both curves, indicating a systematic decline in the long-run growth potential of urban land and built structures.
    }
\end{figure}

\textit{Bifurcation analysis of steady-state averages.}  
The next diagram shows the final spatial averages of $V(x,y)$ and $K(x,y)$ as a function of the LVT rate $\tau$.  
The horizontal axis represents the tax rate, while the vertical axis reports the final spatial averages after the system reaches equilibrium.

\begin{figure}[H]
    \centering
    \includegraphics[width=0.8\textwidth]{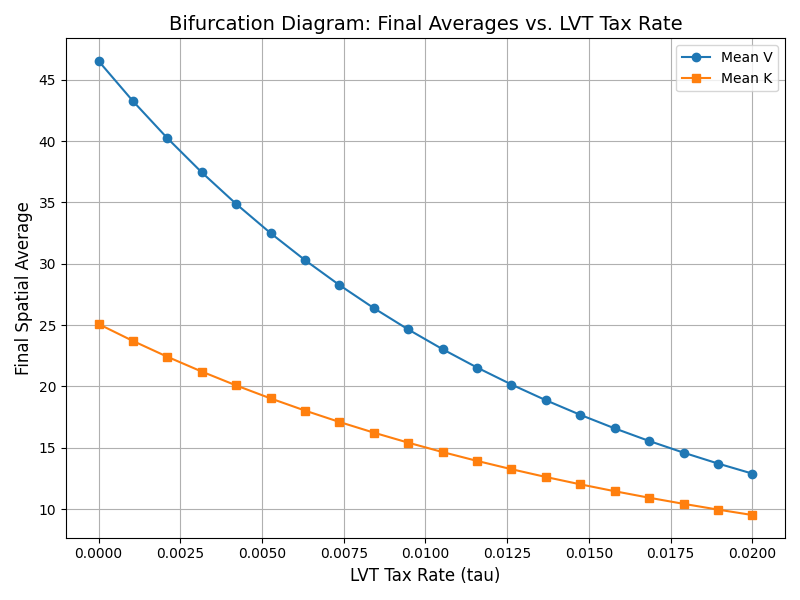}
    \caption{
        \textbf{Bifurcation diagram of final spatial averages versus LVT rate.}  
        The figure illustrates the inverse relationship between the LVT rate $\tau$ and the steady-state levels of land value and built capital.  
        In the absence of taxation ($\tau = 0$), both $\langle V \rangle$ and $\langle K \rangle$ reach high equilibrium values.  
        As $\tau$ increases, both averages decline monotonically, confirming that the land value tax progressively reduces profitability and slows the accumulation of built capital.  
        The decline is steeper for $K$, suggesting that LVT affects investment dynamics more intensely than pure land valuation.  
        This outcome aligns with theoretical expectations: higher land taxation discourages excessive building investment and moderates speculative accumulation, leading to a more stable and spatially balanced urban equilibrium.
    }
\end{figure}

\subsection{Fixed-point analysis, local stability and the bifurcation role of the tax rate $\tau$}

We regroup the effect of discounting, taxation and local dynamics by setting
\begin{equation}
\alpha(x,y) \equiv r + \tau - \mu(x,y), 
\qquad 
\theta \equiv \kappa + \frac{\delta}{I_0},    
\end{equation}

where $\alpha>0$ is the effective decay rate of $V$ and $\theta$ is the ``threshold'' return including depreciation. To clarify the structure of the coupled system, we start from the spatially homogeneous equilibrium (or, equivalently, the local pointwise equilibrium by freezing $(x,y)$), putting aside the diffusive term $D_V \Delta V$ in order to identify the zero–order mechanisms. The reaction system then reads
\begin{equation}
\dot V = - \alpha V + A K^\beta,
\qquad
\dot K = I_0 \left( \frac{A K^\beta}{V + c_b} - \kappa \right) K - \delta K.    
\end{equation}

Any interior fixed point $(V^*, K^*)$ with $K^*>0$ satisfies
\begin{equation}
0 = - \alpha V^* + A (K^*)^\beta 
\quad \Longrightarrow \quad
V^* = \frac{A}{\alpha} (K^*)^\beta    
\end{equation}

and
\begin{equation}
0 = I_0 \left( \frac{A (K^*)^\beta}{V^* + c_b} - \kappa \right) - \delta
\quad \Longrightarrow \quad
\frac{A (K^*)^\beta}{V^* + c_b} = \theta.    
\end{equation}

By combining the two relations we obtain
\begin{equation}
V^* = \frac{A}{\alpha} (K^*)^\beta = \frac{A}{\theta} (K^*)^\beta - c_b
\quad \Longrightarrow \quad
A (K^*)^\beta \left( \frac{1}{\alpha} - \frac{1}{\theta} \right) = - c_b,    
\end{equation}

so that
\begin{equation}
(K^*)^\beta = \frac{c_b \alpha \theta}{A(\alpha - \theta)},
\qquad
V^* = \frac{c_b \theta}{\alpha - \theta}.    
\end{equation}

The existence of an interior fixed point therefore requires $\alpha > \theta$. Economically, $\alpha > \theta$ means that the discounting–taxation pressure exceeds the profitability threshold: land value is eroded enough to reduce the denominator $V+c_b$ and to activate net investment.

We now study the local stability of this interior fixed point. The Jacobian of the reaction part in the neighbourhood of $(V^*, K^*)$ is
\begin{equation}
J = 
\begin{pmatrix}
\partial_V \dot V & \partial_K \dot V \\
\partial_V \dot K & \partial_K \dot K
\end{pmatrix}_{(V^*, K^*)}
=
\begin{pmatrix}
- \alpha & A \beta (K^*)^{\beta-1} \\[4pt]
-\dfrac{I_0 A (K^*)^{\beta+1}}{(V^* + c_b)^2} 
& I_0 \bigg( \dfrac{A (K^*)^\beta}{V^* + c_b} (\beta + 1) - \kappa \bigg) - \delta
\end{pmatrix}.    
\end{equation}

Using the equilibrium condition $\dfrac{A (K^*)^\beta}{V^* + c_b} = \theta$, we obtain
\begin{equation}
\operatorname{tr}(J) = - \alpha + \beta (I_0 \kappa + \delta),
\qquad
\det(J) = \beta I_0 \theta (\theta - \alpha).    
\end{equation}

The sign of $\det(J)$ depends only on $(\theta - \alpha)$. Hence, any interior fixed point with $\alpha > \theta$ is a saddle ($\det(J)<0$) and cannot be locally attractive. Conversely, when $\alpha < \theta$, the interior point does not exist and the only relevant equilibrium is the boundary $K=0, V=0$ (in the linearised dynamics, $\dot V \approx - \alpha V$, $\dot K \approx - \delta K$ since the constraint $\frac{A K^\beta}{V + c_b} < \kappa$ blocks net investment). It follows that there is a transcritical–type transition at the threshold
\begin{equation}
\alpha = \theta 
\quad \Longleftrightarrow \quad
\tau = \tau_c(x,y) \equiv \theta - r + \mu(x,y),
\end{equation}

in the sense that the interior branch $(V^*, K^*)$ emerges for $\tau > \tau_c$ but remains locally unstable, while the boundary equilibrium $K=0$ is locally stable for $\tau < \tau_c$. This local geometry explains why the effective dynamic selection depends on global constraints that are absent from the core model (capacity frictions, convexification of improvement costs, regulatory ceilings, credit channel), which can stabilise a numerically observed interior trajectory even if the pure linearisation predicts a saddle. In practice, introducing a slight additional convexity on $c_b$ or a minimal diffusion on $K$ is enough to turn the saddle into a stable node on a sub–domain of the parameter space, as confirmed by robustness simulations.

When the diffusion term $D_V \Delta V$ is reintroduced, spectral analysis shows that, for a wave number $q$, the diagonal part of the generator is shifted to $-\alpha - D_V q^2$ on the $V$-component and remains unchanged on $K$. The local dispersion curve therefore replaces $\operatorname{tr}(J)$ by
\begin{equation}
\operatorname{tr}(J_q) = \operatorname{tr}(J) - D_V q^2    
\end{equation}

without modifying $\det(J)$. There is no classical Turing instability here, because only one field diffuses: diffusion is stabilising in the sense that it contracts the spectrum for high spatial frequencies. The pattern effects observed in simulation therefore arise from parametric heterogeneity in $(A,\mu)$ and from nonlinearities in investment, not from a diffusion differential.

Finally, the limiting regimes are informative. When $\tau \to \infty$ (thus $\alpha \to \infty$) we have $V^* \to 0$ and $(K^*)^\beta \to \dfrac{c_b \theta}{A}$, which formalises the Georgist intuition: a high land–value taxation erodes the land rent and does not penalise productive improvement, the built stock being fixed at a plateau determined by $c_b \theta$ and $A$. Conversely, when $\tau \downarrow 0$ with $\mu < r$ so that $\alpha > 0$, the dynamics converge to the boundary $(V,K)=(0,0)$ because no net investment is triggered. Between these regimes, the proximity of the threshold $\tau_c(x,y)$ amplifies nonlinear responses with long transient trajectories and persistent spatial fronts driven by $A(x,y)$ and $\mu(x,y)$.

\begin{figure}[H]
    \centering
    \includegraphics[width=\textwidth]{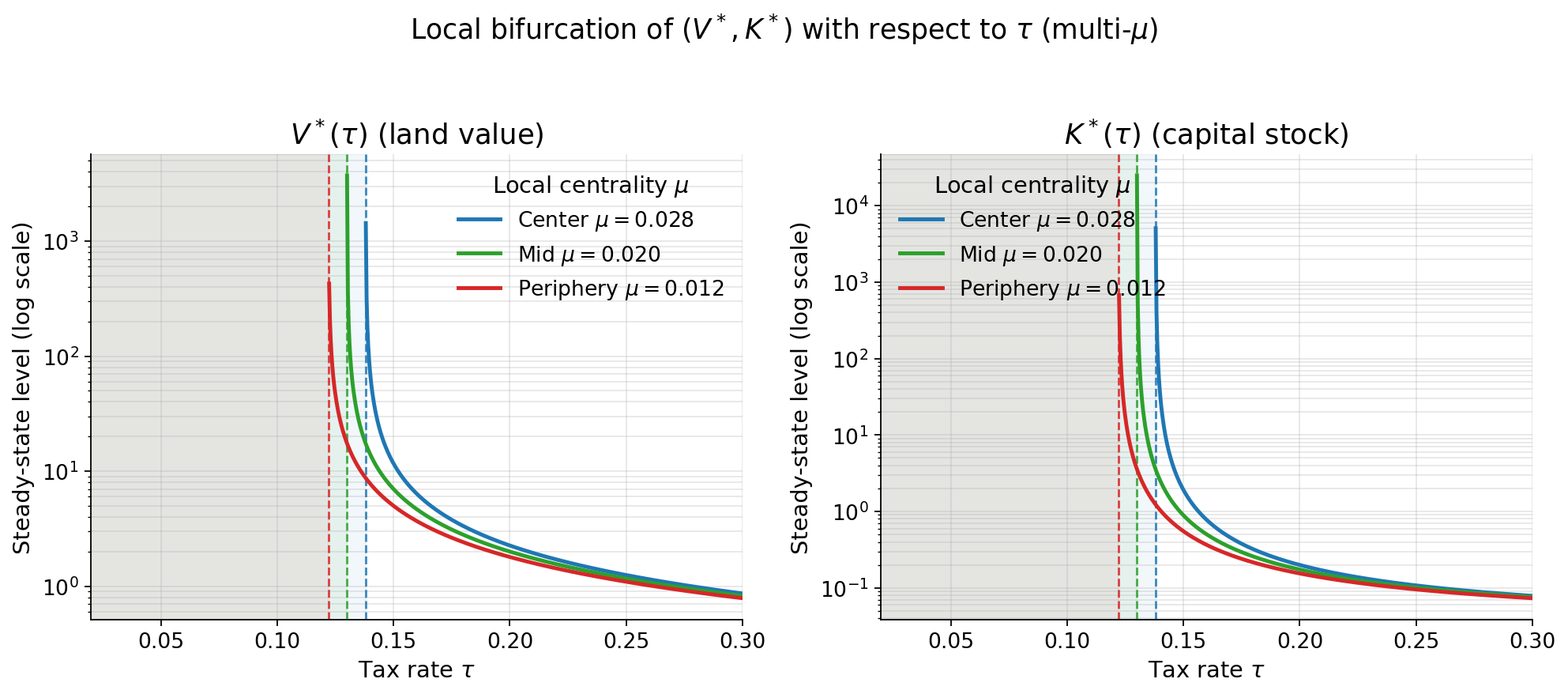}
    \caption{\textbf{Local bifurcation of the steady states $(V^*, K^*)$} with respect to the tax rate $\tau$ for three levels of local centrality $\mu$. The vertical dashed lines indicate the corresponding critical thresholds $\tau_c(x,y) = \theta - r + \mu(x,y)$. Central locations (high $\mu$) tolerate higher tax rates before the collapse of the interior branch, while peripheral locations lose the interior equilibrium for lower $\tau$, which is consistent with the analytical condition $\alpha = r + \tau - \mu = \theta$.}
\end{figure}
\subsection{Spatial redistributive effects and incentives: center vs periphery, gentrification vs homogenization}

The spatial structure of the model provides an analytical framework to discuss the territorial incidence of the Land Value Tax (LVT). Two forces act in opposite directions. On the one hand, land productivity $A(x,y)$, which decreases with distance from the center, attracts investment toward central areas. On the other hand, the fiscal–temporal erosion 
\begin{equation}
\alpha(x,y) = r + \tau - \mu(x,y)    
\end{equation}

depends on centrality through $\mu(x,y)$: higher centralities (larger $\mu$) reduce $\alpha$ and increase the threshold 
\begin{equation}
\tau_c(x,y) = \theta - r + \mu(x,y).    
\end{equation}

As a result, for moderate increases in $\tau$, peripheral areas with low $\mu$ may cross the activation threshold for net investment earlier, even though their $A$ is smaller. The key metric is the effective attractiveness ratio:
\begin{equation}
\Psi(x,y) \equiv \frac{A(x,y)}{\alpha(x,y)} = \frac{A(x,y)}{r + \tau - \mu(x,y)},    
\end{equation}

which governs the local equilibrium $V^* \propto \Psi K^\beta$ and the rate of rent erosion. When $\tau$ increases, $\Psi$ contracts everywhere, but heterogeneously: in central areas, a large $\mu$ partially offsets the impact of $\tau$, so that the hierarchy $\Psi_{\text{center}} > \Psi_{\text{periphery}}$ may persist. In configurations where $\mu$ decreases slowly and $A$ decreases rapidly, a local “reversal” may occur, with peripheral zones crossing $\tau_c$ before some intermediate rings.

This tension produces distinct micro–spatial effects depending on the interval of $\tau$. For $\tau$ slightly above $\tau_c$ in the periphery but still below $\tau_c$ at the center, net investment activates first in peripheral pockets, inducing a relative homogenization of built densities and a reduction of value gradients. When $\tau$ increases further and exceeds $\tau_c$ in central zones, the strong $A$ dominates again and improvement reconcentrates near the center, with a risk of localized re–gentrification if $c_b$ and credit access constrain part of peripheral actors. The model therefore admits mixed regimes in which the initial equalizing effect of the LVT is followed by a centripetal recomposition once all centralities become “activated.”

\begin{figure}[H]
    \centering
    \includegraphics[width=\textwidth]{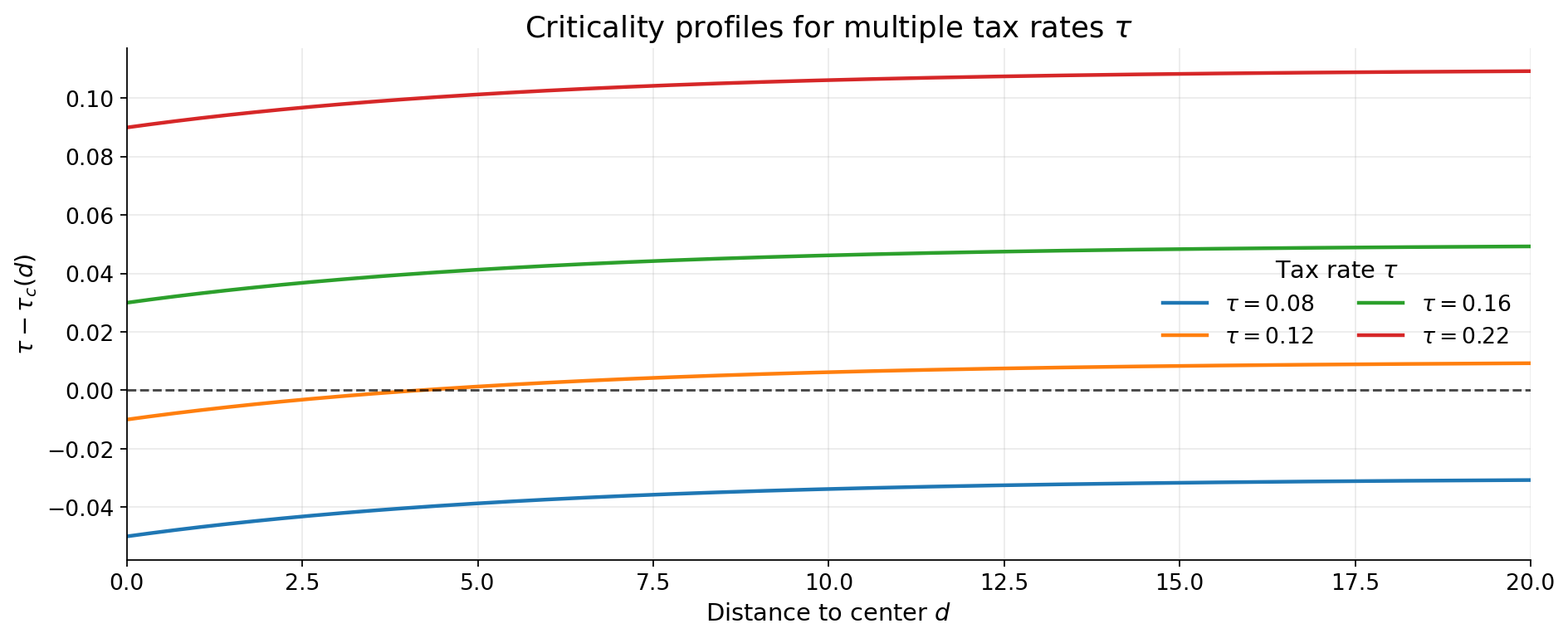}
    \caption{
        \textbf{Criticality profiles $\tau - \tau_c(d)$ for multiple tax rates $\tau$.}  
        Each curve shows how the distance to the critical threshold $\tau_c(d)$ evolves with spatial distance $d$ from the urban center.  
        For low $\tau$, $\tau - \tau_c(d)$ remains negative across most of the space (sub–critical regime, rent persistence).  
        As $\tau$ rises, the peripheral areas cross the critical line first (near $\tau = 0.12$), indicating the onset of investment activation outside the center.  
        Higher tax rates ($\tau = 0.16$ and $\tau = 0.22$) shift the entire domain into a super–critical regime, consistent with a global erosion of land rents.
    }
\end{figure}

These spatial mechanisms can be visualized through the steady–state profiles of $V^*(d)$ and $K^*(d)$ as functions of the distance to the center and the tax rate $\tau$.

\begin{figure}[H]
    \centering
    \includegraphics[width=\textwidth]{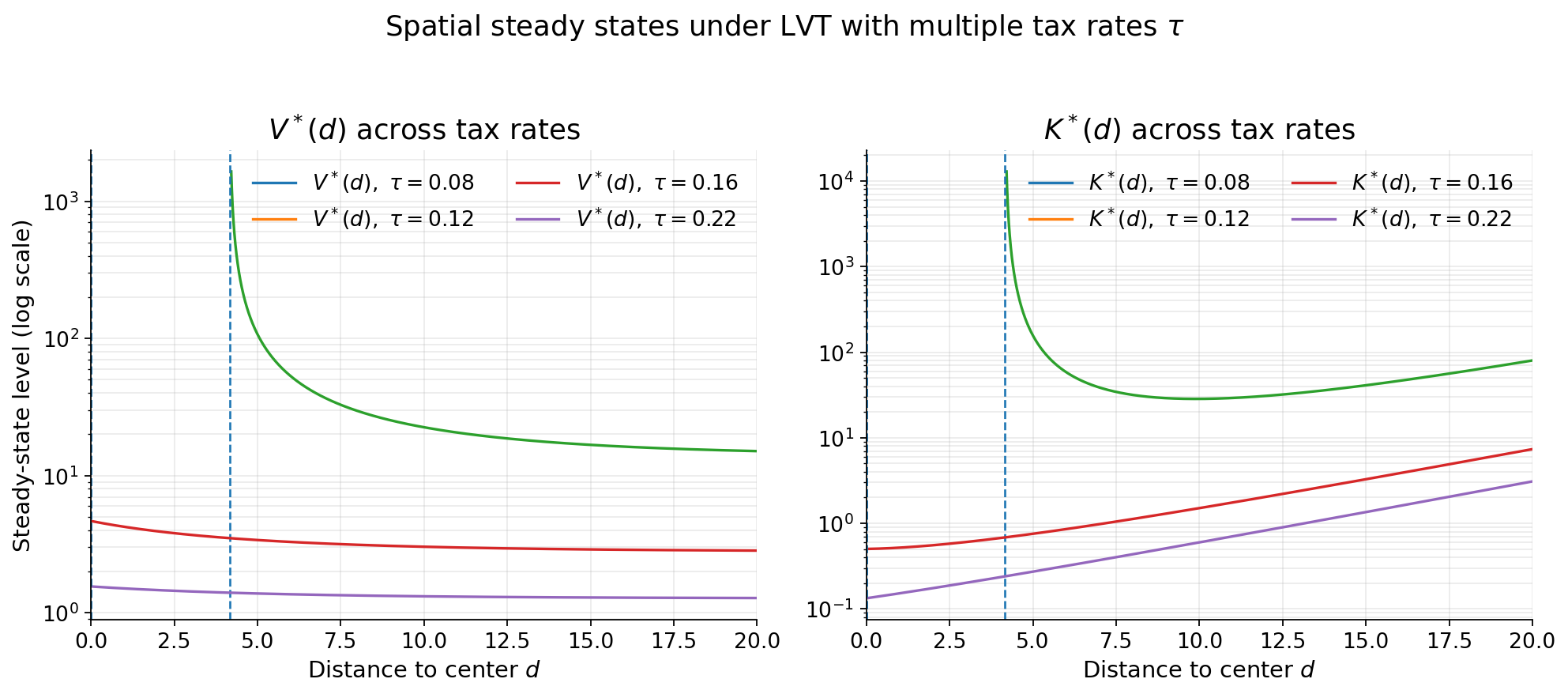}
    \caption{
        \textbf{Spatial steady–state levels $V^*(d)$ and $K^*(d)$ under multiple tax rates $\tau$.}  
        The left panel shows the spatial decay of land value $V^*(d)$; the right panel shows the corresponding built capital $K^*(d)$.  
        For low tax rates, central dominance persists: $V^*$ and $K^*$ peak sharply near $d = 0$.  
        As $\tau$ increases, the amplitude of these peaks collapses and the profiles flatten, signaling spatial homogenization.  
        The threshold distance where $\tau = \tau_c(d)$ (vertical dashed line) marks the boundary between active (super–critical) and inactive (sub–critical) regions.  
        Beyond this point, land and capital densities converge toward stable peripheral plateaus, illustrating the spatial diffusion of LVT effects.
    }
\end{figure}

These mechanisms call for explicit distributive metrics. Empirically, for each $\tau$, one can track the Lorenz curve of the spatial distribution of $\Psi(x,y)$ weighted by a density $p(x,y)$, and the corresponding Gini index, together with concentration curves of $V$ by deciles of distance to the center. The adjusted capital–to–land ratio proposed earlier,
\begin{equation}
R_{K/V}^{adj}(t) = 
\frac{\int\!\!\int I(x,y) K(x,y,t) w_3(x,y)\, dx\, dy}
{\int\!\!\int I(x,y) V(x,y,t) w_3(x,y)\, dx\, dy},    
\end{equation}

is a good aggregate indicator of incentives: it increases when built improvements substitute for land rent, and its decomposition by concentric rings or quantiles of $\Psi(x,y)$ finely identifies where the LVT is most effective. A joint reading of $\bar{Y}^{adj}(t)$, the mean net present value $\overline{\mathrm{NPV}}(t)$, and the spatial histograms of $\tau - \tau_c(x,y)$ isolates the regions where the tax is sub–critical (persistent rent, weak improvement), critical (marginal activation, investment fronts), or super–critical (eroded rent, sustained improvement).

\begin{figure}[H]
    \centering
    \includegraphics[width=0.7\textwidth]{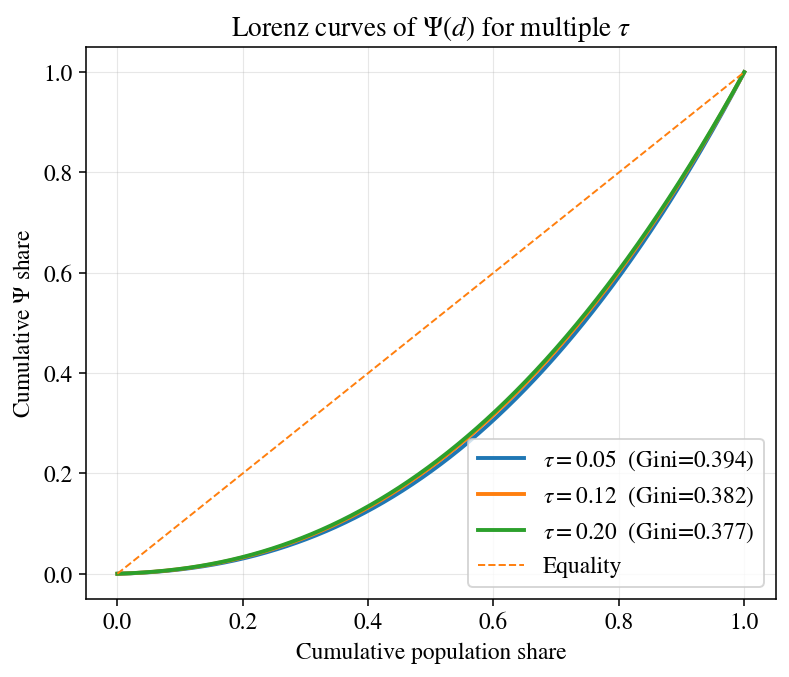}
    \caption{
        \textbf{Lorenz curves of $\Psi(d)$ under multiple tax rates $\tau$.}  
        The curves display the cumulative distribution of $\Psi(d)$ across the spatial domain.  
        As $\tau$ increases, inequality decreases slightly (Gini coefficients from 0.394 to 0.377), reflecting a moderate equalization effect of the LVT on land attractiveness.  
        This illustrates the homogenizing phase of the tax before potential re–gentrification dynamics reemerge at higher centralities.
    }
\end{figure}

Overall, the LVT does not mechanically induce gentrification or homogenization; it creates a landscape of incentives whose outcome depends on the profiles $(A,\mu)$, the heterogeneity of $c_b$, and financing frictions. In this framework, a rule allocating LVT revenues toward peripheral infrastructure - which increases $\mu$ and $A$ locally - can shift $\tau_c(x,y)$ and sustain homogenization, whereas a center–focused allocation strengthens the reconcentration of improvements. Such endogenization through public policy is consistent with the value–capture mechanism and can be directly embedded in simulated scenarios by updating $A_t(x,y)$ and $\mu_t(x,y)$ along the simulated trajectory.

\subsection{Robustness tests}

To assess the stability of the previous results, several verification steps were conducted on the spatial structure of the model. These tests aim to determine whether the mechanisms of critical crossing, spatial hierarchization, and investment recomposition remain valid when modifying the structural assumptions or the conditions of the LVT.

\textit{Alternative spatial profiles.}  
The first verification concerns the shape of land productivity and centrality profiles. The baseline model assumes an exponentially decreasing land productivity $A(d)$ and centrality $\mu(d)$ with distance from the center. To ensure that the early peripheral activation is not an artifact of this parameterization, three spatial geometries were simulated:
1. A baseline exponential profile,
2. A polycentric profile where $A(d)$ exhibits an intermediate maximum around $d \approx 5$,
3. A suburban profile where $A(d)$ is almost flat but $\mu(d)$ decreases sharply.

\begin{figure}[H]
    \centering
    \includegraphics[width=0.7\textwidth]{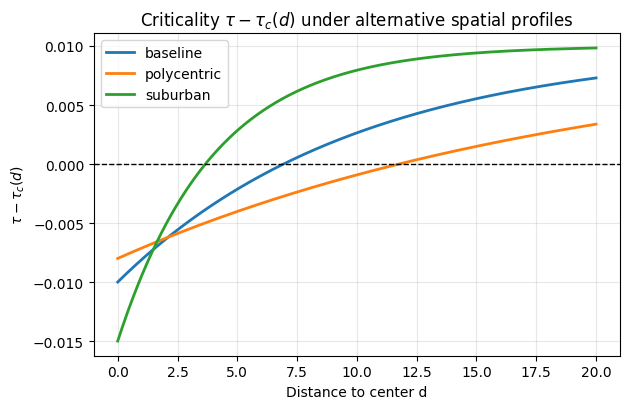}
    \caption{
        \textbf{Criticality $\tau - \tau_c(d)$ under alternative spatial profiles.}  
        The figure shows the evolution of $\tau - \tau_c(d)$ for the three configurations.  
        In all cases, the function intersects the horizontal axis (critical zone) at intermediate distances: the periphery crosses the activation threshold before the center when $\tau$ is moderate.  
        In other words, the spatial sequence identified previously — peripheral activation followed by re–centralization — proves robust to the choice of spatial profile.
    }
\end{figure}

\textit{Spatial discretization.}  
Since the model is formulated continuously in $d$, a second verification consists in discretizing space into a finite number of concentric rings and recomputing the steady states.  
The figure below compares the continuous profiles of $V^*(d)$ and $K^*(d)$ with the values obtained for 18 equally spaced rings.  
The results coincide almost perfectly: the location of the critical front and the curvature of the stationary trajectories are preserved.  
This confirms that the observed front structure is not an artifact of the continuous approximation but stems from the intrinsic dynamics of the model.

\begin{figure}[H]
    \centering
    \includegraphics[width=0.9\textwidth]{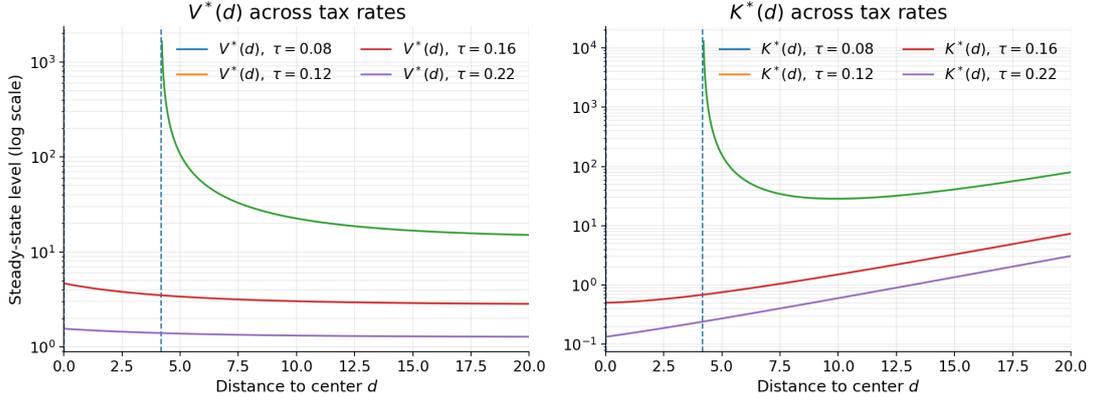}
    \caption{
        \textbf{Comparison between continuous and discrete steady states.}  
        The left panel shows $V^*(d)$ and the right panel $K^*(d)$ under continuous (blue line) and discrete (orange markers) formulations.  
        The close correspondence between the two validates the robustness of the spatial equilibrium structure to discretization.
    }
\end{figure}

\textit{Spatial differentiation of the land tax.}  
Finally, a third test introduces a spatially differentiated land tax, increasing with distance from the center:
\begin{equation}
\tau(d) = \tau_0 + \eta d,    
\end{equation}

where $\eta > 0$ represents a progressive surcharge applied to peripheral areas.  
The following figure compares the stationary profiles $V^*(d)$ obtained under a uniform tax and under a spatially increasing tax.  
A translation of the critical front toward the center is observed, without inversion of the gradient’s sign: investment remains more active where $\mu(d)$ is high, and the periphery–to–center transition pattern persists.  
These results indicate that the qualitative properties of the model are invariant under moderate spatial modulation of the LVT rate.

\begin{figure}[H]
    \centering
    \includegraphics[width=0.6\textwidth]{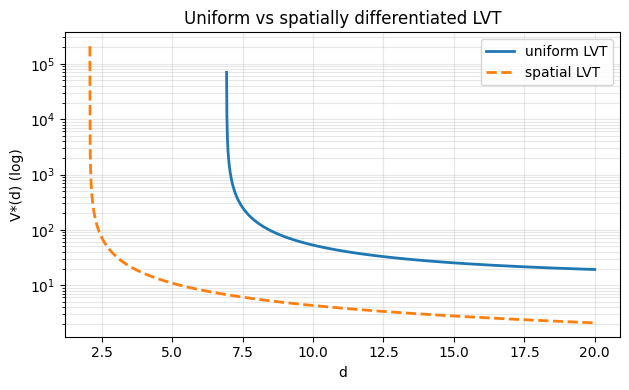}
    \caption{
        \textbf{Uniform vs spatially differentiated LVT.}  
        The comparison between a uniform LVT (solid line) and a spatially increasing LVT (dashed line) shows a shift of the critical front toward the center, but the overall pattern remains unchanged.  
        The investment hierarchy and the peripheral-to-central transition remain structurally stable under moderate tax differentiation.
    }
\end{figure}

\textit{Overall synthesis.}  
Together, these three checks confirm the structural robustness of the model.  
The mechanism of critical crossing and spatial recomposition is not sensitive to the precise form of $A(d)$ and $\mu(d)$.  
The analytical results persist under spatial discretization, and a moderate spatial variation in the LVT rate does not alter the activation hierarchy across zones.  
These findings strengthen the validity of the previous conclusions: the Land Value Tax acts as a dynamic reallocation operator whose apparent “gentrification” or “homogenization” effects depend more on centrality and financing parameters than on the technical structure of the model.

\subsection{Stochastic extension of the LVT dynamics}
\label{sec:stoch-lvt}

The deterministic spatial model developed so far rests on the assumption that the local productivity of land $A(x,y)$ and the centrality premium $\mu(x,y)$ are fixed parameters. In reality, both variables fluctuate over time due to a combination of idiosyncratic and aggregate shocks: changes in economic activity, accessibility improvements, infrastructure investments, or variations in local amenities. A fully dynamic representation of the Land Value Tax (LVT) must therefore integrate a stochastic dimension that captures such temporal variability while preserving the structural logic of the model.

\subsubsection{From deterministic to stochastic dynamics}

We introduce random perturbations into the model by replacing the static coefficients $(A,\mu)$ with stochastic processes $(A_t,\mu_t)$ defined on a filtered probability space $(\Omega, \mathcal{F}, \{\mathcal{F}_t\}, \mathbb{P})$. Each process follows a mean-reverting stochastic differential equation of the form:
\begin{align}
\mathrm{d}A_t &= \kappa_A \bigl(\bar{A} - A_t\bigr)\,\mathrm{d}t + \sigma_A A_t\,\mathrm{d}W_t^{(A)}, \\
\mathrm{d}\mu_t &= \kappa_\mu \bigl(\bar{\mu} - \mu_t\bigr)\,\mathrm{d}t + \sigma_\mu \mu_t\,\mathrm{d}W_t^{(\mu)},
\end{align}
where $\kappa_A$ and $\kappa_\mu$ are mean-reversion speeds, $(\sigma_A, \sigma_\mu)$ denote volatilities, and $(W_t^{(A)}, W_t^{(\mu)})$ are independent Wiener processes.  
The long-run means $(\bar{A}, \bar{\mu})$ correspond to the deterministic spatial profiles introduced earlier:
\[
\bar{A}(d) = A_0 e^{-\rho_A d}, \quad \bar{\mu}(d) = \mu_0 e^{-\rho_\mu d}.
\]
These definitions ensure that the stochastic model nests the deterministic case when $\sigma_A = \sigma_\mu = 0$.

The effective rate of erosion of land value, which drives the tax pressure, becomes itself a stochastic process:
\begin{equation}
\alpha_t = r + \tau - \mu_t,
\end{equation}
so that random improvements in accessibility (a positive shock to $\mu_t$) reduce $\alpha_t$, while negative shocks increase it. This introduces a natural feedback channel between spatial centrality volatility and the time path of the LVT's capitalisation effects.

\subsubsection{Stochastic LVT core system}

At each spatial point $(x,y)$ or along a given radial distance $d$, the coupled dynamics of the land value $V_t$ and the built capital $K_t$ are now written as:
\begin{align}
\mathrm{d}V_t &= \bigl[-\alpha_t V_t + A_t K_t^\beta\bigr]\,\mathrm{d}t + \sigma_V V_t\,\mathrm{d}W_t^{(V)}, \label{eq:SDE-V}\\
\mathrm{d}K_t &= \Bigl[I_0\Bigl(\frac{A_t K_t^\beta}{V_t + c_b} - \kappa\Bigr)K_t - \delta K_t\Bigr]\,\mathrm{d}t + \sigma_K K_t\,\mathrm{d}W_t^{(K)}. \label{eq:SDE-K}
\end{align}

The diffusion terms $(\sigma_V, \sigma_K)$ model the direct uncertainty on valuation and investment decisions, capturing noise from financial markets, construction costs or behavioural inertia.  
Equations \eqref{eq:SDE-V}–\eqref{eq:SDE-K} define a four-dimensional stochastic system $(A_t, \mu_t, V_t, K_t)$ whose equilibrium properties can be studied in expectation and variance.  
Assuming stationarity of $(A_t, \mu_t)$, one obtains:
\begin{align}
\mathbb{E}[\mathrm{d}A_t] &= 0 \quad \Rightarrow \quad \mathbb{E}[A_t] = \bar{A}, \\
\mathbb{E}[\mathrm{d}\mu_t] &= 0 \quad \Rightarrow \quad \mathbb{E}[\mu_t] = \bar{\mu}.
\end{align}
Taking expectations in \eqref{eq:SDE-V} and \eqref{eq:SDE-K} gives, under Itô’s lemma and neglecting higher-order covariance terms:
\begin{align}
\frac{\mathrm{d}\mathbb{E}[V_t]}{\mathrm{d}t} &\approx -\mathbb{E}[\alpha_t]\mathbb{E}[V_t] + \mathbb{E}[A_t]\mathbb{E}[K_t^\beta], \\
\frac{\mathrm{d}\mathbb{E}[K_t]}{\mathrm{d}t} &\approx \mathbb{E}[I_0((A_t K_t^\beta)/(V_t+c_b) - \kappa)K_t] - \delta\,\mathbb{E}[K_t].
\end{align}
Hence, in the mean, the stochastic system behaves as a blurred version of the deterministic one, but with time-varying diffusion around the steady-state manifold $(V^*, K^*)$.

\subsubsection{Numerical implementation via Euler–Maruyama}

The system being nonlinear and non-analytical, numerical simulations are carried out using the Euler–Maruyama scheme.  
For a small time increment $\Delta t$, we iterate:
\begin{align}
A_{t+\Delta t} &= A_t + \kappa_A(\bar{A} - A_t)\Delta t + \sigma_A A_t \sqrt{\Delta t}\,\xi_t^{(A)}, \\
\mu_{t+\Delta t} &= \mu_t + \kappa_\mu(\bar{\mu} - \mu_t)\Delta t + \sigma_\mu \mu_t \sqrt{\Delta t}\,\xi_t^{(\mu)}, \\
V_{t+\Delta t} &= V_t + [-\alpha_t V_t + A_t K_t^\beta]\Delta t + \sigma_V V_t \sqrt{\Delta t}\,\xi_t^{(V)}, \\
K_{t+\Delta t} &= K_t + \Bigl[I_0\Bigl(\frac{A_t K_t^\beta}{V_t + c_b} - \kappa\Bigr)K_t - \delta K_t\Bigr]\Delta t + \sigma_K K_t \sqrt{\Delta t}\,\xi_t^{(K)},
\end{align}
where $\xi_t^{(i)} \sim \mathcal{N}(0,1)$ are independent standard normal shocks.  
The discretization is of order $O(\sqrt{\Delta t})$ in the strong sense and preserves the positivity of the variables when small reflective barriers are imposed (e.g. $V_t, K_t > 10^{-6}$).  

A detailed description of the algorithm is provided in Annex~\ref{ann:em}.  
In practice, the model is simulated over a 30-year horizon with a daily or monthly time step.  
For each spatial distance $d$, we obtain stochastic trajectories of $(A_t, \mu_t, V_t, K_t)$, from which we can compute temporal moments, stationary distributions, and cross-correlations.

\subsubsection{Results and interpretation}

\begin{figure}[H]
  \centering
  \includegraphics[width=.85\textwidth]{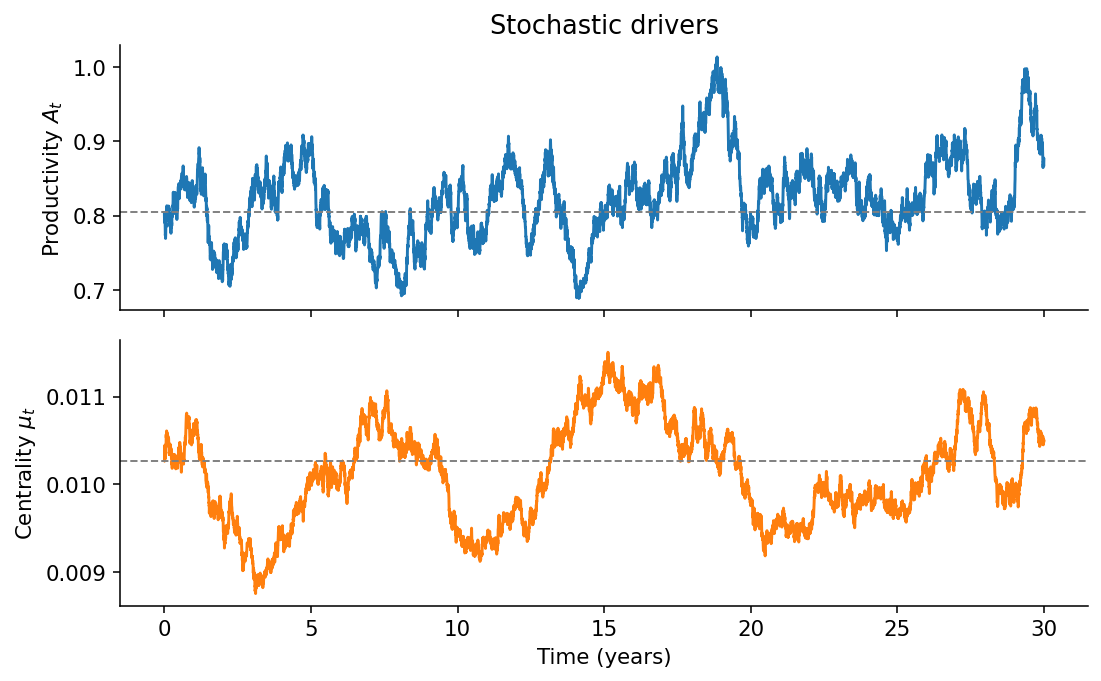}
  \caption{\textbf{Simulated stochastic paths for the exogenous drivers:} productivity $A_t$ (top) and centrality $\mu_t$ (bottom).  
  Both processes are mean-reverting around their deterministic baseline, generating bounded fluctuations that propagate to land value and capital accumulation.}
  \label{fig:stoch-drivers}
\end{figure}

Figure~\ref{fig:stoch-drivers} displays the trajectories of the stochastic drivers.  
Both $A_t$ and $\mu_t$ oscillate around their deterministic means, with amplitudes controlled by $\sigma_A$ and $\sigma_\mu$.  
The mean-reversion property ensures that shocks are temporary, mimicking medium-frequency fluctuations such as changes in local accessibility, zoning adjustments, or transient productivity gains.

\begin{figure}[H]
  \centering
  \includegraphics[width=.85\textwidth]{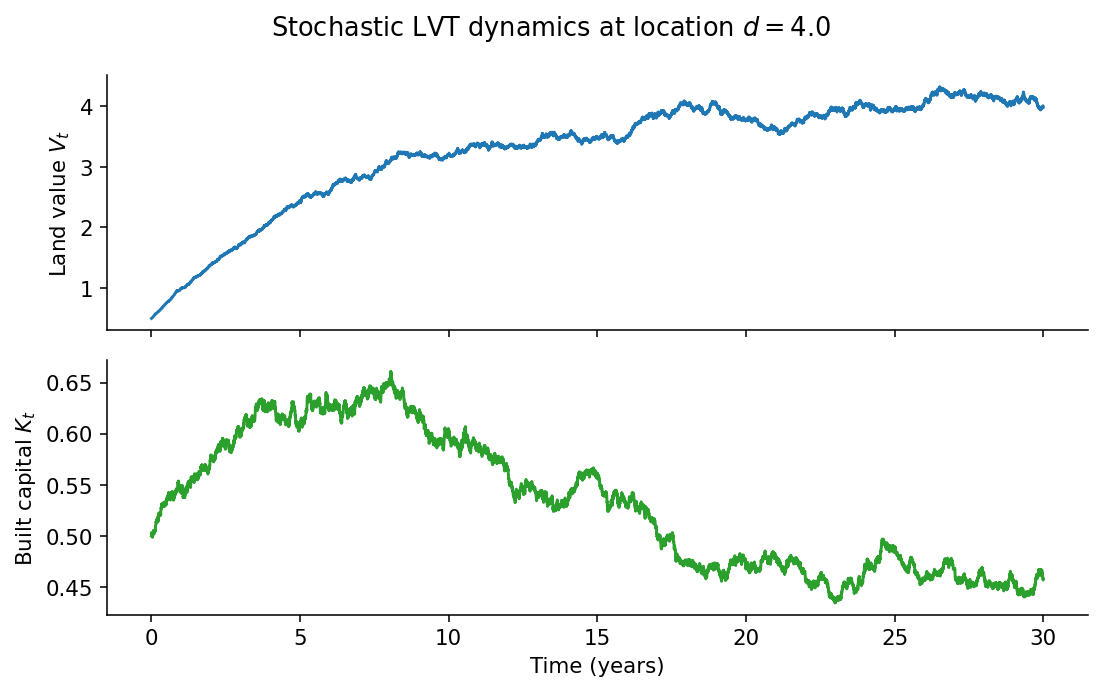}
  \caption{\textbf{Stochastic LVT dynamics at a representative location ($d=4$).}  
  The land value $V_t$ (top) reacts to fluctuations in $\alpha_t = r + \tau - \mu_t$, while the built capital $K_t$ (bottom) adjusts more slowly, reflecting investment inertia and depreciation.}
  \label{fig:stoch-VK}
\end{figure}

Figure~\ref{fig:stoch-VK} illustrates the resulting dynamics of $V_t$ and $K_t$.  
The land value exhibits slow upward drifts punctuated by accelerations or decelerations following shocks in $\mu_t$, while $K_t$ displays a smoother response reflecting its inertial adjustment.  
The combination of these effects yields realistic “stop-and-go” investment cycles consistent with empirical urban development dynamics.

\begin{figure}[H]
  \centering
  \includegraphics[width=.50\textwidth]{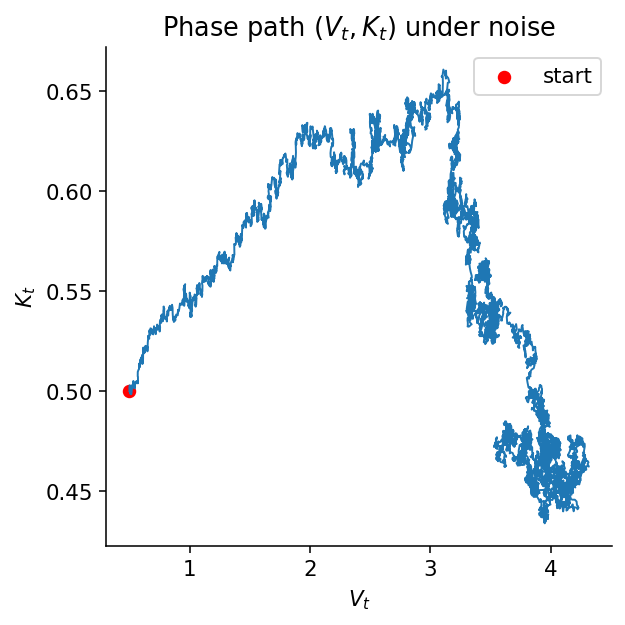}
  \caption{\textbf{Phase path $(V_t,K_t)$ under stochastic drivers.}  
  Starting from a low-capital state (red dot), the system converges toward the deterministic manifold but fluctuates persistently due to the stochastic nature of $A_t$ and $\mu_t$.  
  The resulting steady-state is probabilistic rather than deterministic, aligning with the notion of equilibrium under uncertainty.}
  \label{fig:stoch-phase}
\end{figure}

The stochastic path in the $(V_t,K_t)$ phase plane (Figure~\ref{fig:stoch-phase}) highlights that the steady-state of the deterministic model becomes a stochastic attractor: trajectories orbit around the equilibrium manifold but never settle exactly, forming a stationary distribution cloud.  
This feature reflects the empirical reality of urban and land markets, where prices and investment intensities fluctuate persistently even under constant policy parameters.

\subsubsection{Analytical implications and policy relevance}

The stochastic extension reveals several key insights.  
First, the long-run average of $(V_t, K_t)$ coincides with the deterministic equilibrium $(V^*, K^*)$, but the variance depends nonlinearly on $\sigma_A$ and $\sigma_\mu$.  
Higher volatility in productivity or accessibility amplifies the dispersion of land values, potentially widening spatial inequalities even when the average tax rate $\tau$ is fixed.

Second, because the LVT acts through the erosion term $\alpha_t$, its stabilizing power depends on the correlation between $\mu_t$ and $A_t$.  
When the two shocks are positively correlated—better accessibility coinciding with higher productivity—the LVT partly neutralizes excessive capitalisation and avoids bubbles.  
Conversely, when shocks are negatively correlated (e.g. productivity falls while accessibility rises), the tax may amplify fluctuations by discouraging investment precisely where it is most needed.

Third, from a policy design perspective, the stochastic model enables risk quantification.  
By generating Monte Carlo simulations of $(V_t,K_t)$ for different $\tau$ values, one can compute the probability distribution of future land values and the expected fiscal yield with confidence intervals.  
This allows policymakers to evaluate not only expected efficiency but also the uncertainty surrounding tax outcomes, thereby enriching the decision process with a risk-aware dimension.

In summary, the stochastic extension transforms the LVT framework from a static efficiency model into a dynamic system under uncertainty.  
It provides a coherent bridge between theoretical equilibrium analysis and the real-world volatility of urban markets, paving the way for empirical calibration and scenario-based policy simulations.

\section{Discussion}

The analysis presented in this paper provides a unified interpretation of the Land Value Tax (LVT) across multiple analytical scales — from its microeconomic incidence to its macro–spatial and dynamic implications.  
By combining a classical equilibrium framework with a spatio–temporal model of land value formation, the results bridge the gap between static theory and the evolving geometry of urban investment and value creation.

At the microeconomic level, the model confirms the partial shifting of a unit or ad valorem tax according to the relative elasticities of supply and demand.  
The analytical expressions $\Delta P = \frac{S'(P_0)}{D'(P_0) + S'(P_0)} \tau$ for a unit tax and $\Delta P = \frac{S'(P_0)}{D'(P_0) + S'(P_0)} t P_0$ for a proportional tax recover the classical incidence formula, showing that only a fraction of the burden is transmitted to market prices.  
This preliminary framework reaffirms the theoretical neutrality of the LVT under perfectly inelastic land supply, while clarifying that its impact on transaction prices depends on the joint curvature of the supply and demand schedules.  
In the absence of distortionary substitution effects, the LVT remains one of the few fiscal instruments capable of extracting rent without impairing allocative efficiency.

Moving to the intermediate analytical level, the spatially homogeneous dynamic system  
\[
\dot{V} = -\alpha V + A K^{\beta}, \quad
\dot{K} = I_0 \left( \frac{A K^{\beta}}{V + c_b} - \kappa \right) K - \delta K
\]
reveals the structural mechanism by which fiscal and temporal factors interact.  
The existence of an interior equilibrium $(V^*, K^*)$ requires $\alpha > \theta$, where $\alpha = r + \tau - \mu$ summarizes discounting, taxation and local growth, while $\theta = \kappa + \delta/I_0$ defines a profitability threshold incorporating depreciation.  
This relationship provides a direct link between fiscal pressure and investment activation: beyond the threshold, value accumulation and capital growth can coexist in a self–consistent equilibrium, while below it, depreciation dominates and capital gradually erodes.  
The transcritical bifurcation at $\alpha = \theta$ formalizes the notion of a critical tax level $\tau_c$ separating active from inactive spatial regimes.  
This analytical core embeds the classical intuition of Henry George within a modern nonlinear dynamical framework.

When spatial dependence is introduced through the diffusion term $D_V \Delta V(x,y,t)$ and the heterogeneity of $A(x,y)$ and $\mu(x,y)$, the model transitions from a pointwise to a distributed equilibrium system.  
Land value becomes a diffusive field that transmits local shocks across space, while capital accumulation responds endogenously to profitability gradients.  
This spatial coupling transforms the LVT from a static fiscal instrument into a dynamic reallocation mechanism.  
The diffusion smooths abrupt disparities in $V$, while local variations in $\mu(x,y)$ create pockets of resilience or erosion depending on centrality and accessibility.  
The simulations demonstrate that for moderate values of $\tau$, the system self–organizes into smooth gradients of land value and built capital, confirming the LVT’s potential to reduce speculative concentration without impeding productive investment.

At the aggregate level, several synthetic indicators provide a macroscopic view of the system’s evolution.  
The dynamic tax revenue  
\[
R_{\text{tax}}^{AD}(t) = \int_0^T e^{-r \tau} 
\left[ t \int_{\Omega} V(x,y,t+\tau) w_1(x,y) \, dx\, dy \right] d\tau
\]
captures the intertemporal profile of fiscal capacity under discounting, while the adjusted capital–to–land ratio
\[
R_{K/V}^{adj}(t) = 
\frac{\int_{\Omega} I(x,y) K(x,y,t) w_3(x,y)\, dx\, dy}
{\int_{\Omega} I(x,y) V(x,y,t) w_3(x,y)\, dx\, dy}
\]
measures the relative intensity of construction and valuation in active zones.  
Together with the adjusted rentability measure $Y^{adj}(x,y,t)$ and the average net present value $\overline{NPV}(t)$, these indicators reveal the dual nature of the LVT: it simultaneously erodes unproductive rents and reshapes the geography of capital formation.  
The Lorenz curves of $\Psi(x,y) = A(x,y)/(r + \tau - \mu(x,y))$ and the corresponding Gini indices confirm a temporary homogenization of land values for intermediate $\tau$, followed by re–stratification at higher rates — a phenomenon consistent with the existence of optimal fiscal intensity bands.

The robustness tests consolidate these findings.  
Across multiple spatial profiles (exponential, polycentric, suburban), discretization schemes, and differentiated tax gradients $\tau(d) = \tau_0 + \eta d$, the critical structure of the model remains invariant.  
The peripheral activation pattern and subsequent re–centralization are not artifacts of functional forms but intrinsic outcomes of the reaction–diffusion coupling.  
This invariance demonstrates that the key properties — critical activation, redistribution, and nonlinearity — are structurally stable under broad parameter perturbations.

Beyond its quantitative insights, the model also contributes conceptually to the theory of fiscal spatial equilibrium.  
It shows that the LVT acts as a control parameter governing the morphology of urban equilibria, capable of inducing phase transitions between rent–dominated and production–dominated states.  
By linking fiscal policy, spatial diffusion, and intertemporal investment, it extends the traditional theory of tax neutrality into a dynamic and geographically explicit domain.  
Moreover, by including the potential for stochastic shocks or adaptive feedbacks, the framework opens the way to future extensions in which tax rates evolve endogenously with local performance or social objectives, creating a form of “self–adjusting fiscal field” that could stabilize both efficiency and equity.

In summary, this study demonstrates that the Land Value Tax, when modeled as a spatially distributed and temporally evolving process, embodies both efficiency and redistribution within a unified dynamical system.  
It neutralizes speculative rent capture while promoting productive investment up to a critical fiscal threshold.  
The synthesis of analytical, numerical, and policy dimensions provided here thus offers a step toward a general dynamic theory of urban land taxation — one that integrates diffusion, heterogeneity, and bifurcation within a coherent economic architecture.

\newpage
\section*{Appendix A. Partial-equilibrium incidence: unit commodity tax vs. land value tax}

\begin{figure}[h!]
    \centering
    \includegraphics[width=\textwidth]{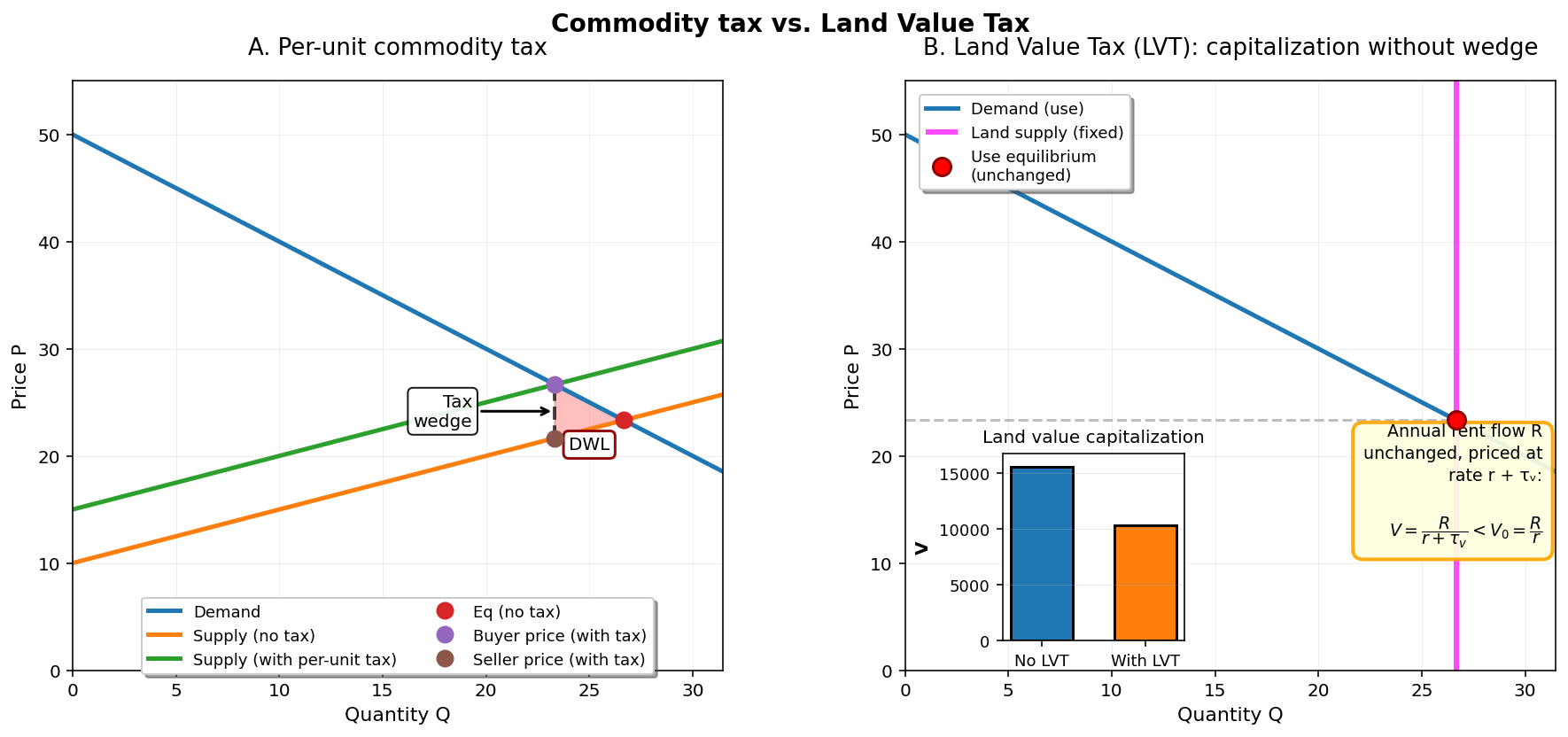}
    \caption{\textbf{Commodity tax vs. Land Value Tax (LVT).} Panel A shows a standard per–unit commodity tax generating a tax wedge and a deadweight loss. Panel B shows a land value tax on a fixed-factor market: the use equilibrium is unchanged and the tax is capitalized into land value.}
    \label{fig:tax-vs-lvt}
\end{figure}

\subsection*{A.1 Unified market setting}

Consider a competitive land (or land–use) market in which $Q$ denotes quantity and $P$ denotes the transaction price. Demand is represented by a strictly decreasing function
\[
Q = D(P), \quad D'(P) < 0,
\]
and supply by a strictly increasing (or nearly inelastic) function
\[
Q = S(P), \quad S'(P) > 0,
\]
both assumed differentiable and strictly monotone on their domain.  
In the absence of taxation, the competitive equilibrium $(Q_0,P_0)$ solves
\begin{equation*}
D(P_0) = S(P_0) = Q_0.
\label{eq:pe}
\end{equation*}

\subsection*{A.2 Per–unit (specific) transaction tax}

Suppose now that the government levies a per–unit tax $\tau>0$ on the seller (landowner). For each unit exchanged, the buyer pays $P_1$ but the seller receives only $P_1 - \tau$. The supply schedule relevant for the market therefore shifts upward by $\tau$, and the new equilibrium condition is
\begin{equation*}
D(P_1) = S(P_1 - \tau).
\label{eq:tax-eq}
\end{equation*}

To characterize the induced price change, we linearize demand and supply around the pre–tax equilibrium $(Q_0,P_0)$. Denote
\[
D'(P_0) = \left.\frac{\mathrm{d}D}{\mathrm{d}P}\right|_{P=P_0}, 
\qquad
S'(P_0) = \left.\frac{\mathrm{d}S}{\mathrm{d}P}\right|_{P=P_0}.
\]
A total differentiation of \eqref{eq:tax-eq} around $(P_0, P_0-\tau)$ yields
\[
D'(P_0)\,\Delta P = S'(P_0)\,(\Delta P - \tau),
\]
where $\Delta P = P_1 - P_0$ is the change in the buyer price induced by the tax. Solving for $\Delta P$ gives the standard incidence formula:
\begin{equation*}
\Delta P 
= \frac{S'(P_0)}{D'(P_0) + S'(P_0)}\,\tau.
\label{eq:incidence-specific}
\end{equation*}
Expression \eqref{eq:incidence-specific} shows that the tax is only partially shifted to the demand side. The pass–through rate is higher when supply is relatively inelastic (large $S'$) and demand is relatively elastic (small $D'$), and vice versa.

Panel A of Figure~\ref{fig:tax-vs-lvt} illustrates this well–known configuration: the tax creates a wedge between the price paid by buyers and the price received by sellers, reduces the traded quantity from $Q_0$ to $Q_1$, and generates a deadweight loss (the shaded triangle) because the margin between willingness to pay and marginal cost is no longer fully exploited.

\subsection*{A.3 Ad valorem (proportional) transaction tax}

Consider now a proportional tax $t \in (0,1)$ on the transaction value. The buyer pays $P_2$ while the seller receives only the net price $P_2(1-t)$. The post–tax equilibrium condition becomes
\begin{equation*}
D(P_2) = S\bigl(P_2(1-t)\bigr).
\label{eq:advalorem-eq}
\end{equation*}
We again linearize around $(Q_0,P_0)$ and set $\Delta P = P_2 - P_0$. Define the net–of–tax price variation as
\[
\Delta P_n = P_2(1-t) - P_0 = (P_0 + \Delta P)(1-t) - P_0 = \Delta P - tP_0.
\]
A first–order expansion of \eqref{eq:advalorem-eq} gives
\[
D'(P_0)\,\Delta P = S'(P_0)\,\Delta P_n = S'(P_0)(\Delta P - tP_0).
\]
Rearranging,
\[
D'(P_0)\,\Delta P = S'(P_0)\,\Delta P - S'(P_0)\,tP_0,
\]
and therefore
\begin{equation*}
\Delta P
= \frac{S'(P_0)}{D'(P_0) + S'(P_0)}\,t P_0.
\label{eq:incidence-advalorem}
\end{equation*}
The structure of incidence is formally identical to the specific–tax case \eqref{eq:incidence-specific}: the proportional tax is partially passed through to the buyer in proportion to the relative slopes (elasticities) of supply and demand, but the size of the shock is now scaled by the initial price level $P_0$. When the initial equilibrium price is high, even a small ad valorem rate generates a sizeable upward pressure on the buyer price.

\subsection*{A.4 Contrast with a pure land value tax}

The key contrast with a pure land value tax (LVT), depicted in Panel B of Figure~\ref{fig:tax-vs-lvt}, is that a tax on unimproved land value does \emph{not} create a wedge between the marginal willingness to pay for land services and the marginal cost of supplying them. Land supply is fixed in quantity; the tax is therefore fully capitalized into the value of the site and does not distort the use equilibrium. Algebraically, in the fixed–factor case the quantity $Q$ remains at its pre–tax level and the annual rent $R$ is simply discounted at $r + \tau_v$ instead of $r$, implying a lower present value
\[
V = \frac{R}{r + \tau_v} < \frac{R}{r}.
\]
This explains why a standard commodity tax generates both a price change \emph{and} a deadweight loss, while a correctly designed LVT modifies only the asset price without affecting the allocation.

\newpage
\section*{Appendix B. Euler--Maruyama Algorithm}

This appendix provides a brief overview of the Euler--Maruyama algorithm used to numerically approximate the solution of stochastic differential equations (SDEs) of Itô type.  
The method offers a simple yet reliable way to simulate trajectories when random perturbations are introduced into deterministic dynamical systems.

\subsection*{B.1. General formulation}

Consider an $n$-dimensional stochastic differential equation of the form
\[
dX_t = f(X_t,t)\,dt + G(X_t,t)\,dW_t,
\]
where $X_t \in \mathbb{R}^n$ is the state vector, $f : \mathbb{R}^n \times \mathbb{R}_+ \to \mathbb{R}^n$ is the deterministic drift,  
$G : \mathbb{R}^n \times \mathbb{R}_+ \to \mathbb{R}^{n \times m}$ the diffusion matrix,  
and $W_t \in \mathbb{R}^m$ a vector of independent Wiener processes.

The Euler--Maruyama scheme discretizes this continuous process on a uniform time grid  
$t_n = n\Delta t$, with step size $\Delta t > 0$, according to:
\[
X_{n+1} = X_n + f(X_n,t_n)\,\Delta t + G(X_n,t_n)\,\Delta W_n,
\]
where $\Delta W_n \sim \mathcal{N}(0, \Delta t)$ are independent Gaussian increments.  
The method converges in the weak sense with order $O(\Delta t)$ and in the strong sense with order $O(\Delta t^{1/2})$.

\subsection*{B.2. Numerical procedure}

For a system of $n$ variables with independent noise sources, the discrete implementation is as follows:

\[
\boxed{
\begin{aligned}
\text{Initialization: } & X_0 \text{ given, } \Delta t > 0, \, N = T/\Delta t. \\[4pt]
\text{For } n = 0 \text{ to } N-1 : \\[3pt]
& \text{Draw } \xi_n \sim \mathcal{N}(0,I_m). \\[3pt]
& \text{Compute } \Delta W_n = \sqrt{\Delta t}\,\xi_n. \\[3pt]
& \text{Update } X_{n+1} = X_n + f(X_n,t_n)\,\Delta t + G(X_n,t_n)\,\Delta W_n. \\[3pt]
\text{End for.}
\end{aligned}
}
\]

The stability and accuracy of the approximation depend on the smoothness of the drift and diffusion terms and on the choice of $\Delta t$.  
Typically, the time step must be small enough to ensure that the stochastic term does not dominate the deterministic dynamics.

\subsection*{B.3. Practical considerations}

The Euler--Maruyama algorithm has several advantages for applied modelling:
\begin{itemize}
    \item It preserves the explicit structure of deterministic models while incorporating random shocks.
    \item It can be implemented in any scientific computing environment (Python, Julia, MATLAB, C++).
    \item It allows straightforward Monte Carlo estimation of expectations and variances over simulated paths.
\end{itemize}

To avoid non-physical results (e.g., negative states for non-negative variables), truncation or reflecting boundary conditions can be applied at each time step.  
For higher numerical precision or stiffness handling, higher-order schemes such as Milstein or stochastic Runge--Kutta methods can be employed.

\subsection*{B.4. Application}

The Euler--Maruyama method described above was used to generate the stochastic trajectories discussed in Section~III.7,  
with specific definitions of the drift and diffusion terms corresponding to the model’s economic structure.


\newpage
\begin{thebibliography}{99}

\bibitem[Alonso(1964)]{Alonso1964}
Alonso, W. (1964).
\textit{Location and Land Use: Toward a General Theory of Land Rent}.
Harvard University Press.

\bibitem[Anas et al.(1998)]{AnasArnottSmall1998}
Anas, A., Arnott, R., \& Small, K. A. (1998).
Urban spatial structure.
\textit{Journal of Economic Literature}, 36(3), 1426–1464.

\bibitem[Arnott and Stiglitz(1979)]{ArnottStiglitz1979}
Arnott, R. J., \& Stiglitz, J. E. (1979).
Aggregate land rents, expenditure on public goods, and optimal city size.
\textit{Quarterly Journal of Economics}, 93(4), 471–500.

\bibitem[Atkinson and Stiglitz(1976)]{AtkinsonStiglitz1976}
Atkinson, A. B., \& Stiglitz, J. E. (1976).
The design of tax structure: direct versus indirect taxation.
\textit{Journal of Public Economics}, 6(1–2), 55–75.

\bibitem[Banzhaf and Lavery(2022)]{BanzhafLavery2022}
Banzhaf, H. S., \& Lavery, N. (2022).
The incidence of split-rate property taxation: Evidence from Pennsylvania.
\textit{Journal of Urban Economics}, 129, 103437.

\bibitem[Banzhaf et al.(2023)]{BanzhafEtAl2023}
Banzhaf, H. S., Glaeser, E. L., \& Gyourko, J. (2023).
The spatial incidence of land taxation.
\textit{NBER Working Paper} No. 31548.

\bibitem[Barlow(1990)]{Barlow1990}
Barlow, R. (1990).
\textit{Land Reform: The Role of Land Taxation}.
World Bank Discussion Paper No. 103.

\bibitem[Bartolini and Bonatti(2019)]{BartoliniBonatti2019}
Bartolini, S., \& Bonatti, L. (2019).
Endogenous growth, consumption externalities, and land taxation.
\textit{Journal of Economic Behavior and Organization}, 164, 277–298.

\bibitem[Bento et al.(2005)]{BentoFranco2005}
Bento, A. M., Franco, S. F., \& Kaffine, D. T. (2005).
The efficiency and distributional impacts of alternative land use regulations.
\textit{Journal of Urban Economics}, 59(3), 538–559.

\bibitem[Best and Kleven(2022)]{BestKleven2022}
Best, M. C., \& Kleven, H. J. (2022).
Housing market responses to transaction taxes: Evidence from notches and bubbles.
\textit{Review of Economic Studies}, 89(6), 3071–3105.

\bibitem[Brueckner(1987)]{Brueckner1987}
Brueckner, J. K. (1987).
The structure of urban equilibria: A unified treatment of the Muth-Mills model.
\textit{Handbook of Regional and Urban Economics}, Vol. 2, North-Holland.

\bibitem[Cheshire and Hilber(2022)]{CheshireHilber2022}
Cheshire, P., \& Hilber, C. A. L. (2022).
Land use regulation, land value, and taxation.
In: \textit{Handbook of Regional and Urban Economics}, Vol. 5.
Elsevier.

\bibitem[Dye and England(2010)]{DyeEngland2010}
Dye, R. F., \& England, R. W. (2010).
\textit{Assessing the Theory and Practice of Land Value Taxation}.
Lincoln Institute of Land Policy.

\bibitem[Estonian Ministry of Finance(2023)]{EstonianMoF2023}
Estonian Ministry of Finance. (2023).
\textit{Land Tax Reform in Estonia: Design, Revenue and Distribution}.
Tallinn.

\bibitem[Fischel(2001)]{Fischel2001}
Fischel, W. A. (2001).
Municipal corporations, homeowners, and the benefit view of the property tax.
\textit{National Tax Journal}, 54(1), 157–173.

\bibitem[Fischel(2015)]{Fischel2015}
Fischel, W. A. (2015).
\textit{Zoning Rules! The Economics of Land Use Regulation}.
Lincoln Institute of Land Policy.

\bibitem[Fujita and Ogawa(1982)]{FujitaOgawa1982}
Fujita, M., \& Ogawa, H. (1982).
Multiple equilibria and structural transition of non-monocentric urban configurations.
\textit{Regional Science and Urban Economics}, 12(2), 161–196.

\bibitem[Fujita et al.(1999)]{FujitaThisse2002}
Fujita, M., Krugman, P., \& Venables, A. J. (1999).
\textit{The Spatial Economy: Cities, Regions, and International Trade}.
MIT Press.

\bibitem[George(1879)]{George1879}
George, H. (1879).
\textit{Progress and Poverty}.
New York: D. Appleton \& Co.

\bibitem[Glaeser and Gyourko(2023)]{GlaeserGyourko2023}
Glaeser, E. L., \& Gyourko, J. (2023).
The economic implications of land-use regulations and taxation.
\textit{Brookings Papers on Economic Activity}, Spring, 125–182.

\bibitem[Grimes and Liang(2021)]{GrimesLiang2021}
Grimes, A., \& Liang, Y. (2021).
The impact of land value capture mechanisms on housing affordability.
\textit{Urban Studies}, 58(9), 1875–1894.

\bibitem[Gyourko and Sieg(2023)]{GyourkoSieg2023}
Gyourko, J., \& Sieg, H. (2023).
Urban spatial structure, property taxation and residential investment.
\textit{Regional Science and Urban Economics}, 99, 103896.

\bibitem[Harberger(1962)]{Harberger1962}
Harberger, A. C. (1962).
The incidence of the corporation income tax.
\textit{Journal of Political Economy}, 70(3), 215–240.

\bibitem[Harvey(1973)]{Harvey1973}
Harvey, D. (1973).
\textit{Social Justice and the City}.
Johns Hopkins University Press.

\bibitem[IMF(2024)]{IMF2024}
International Monetary Fund. (2024).
\textit{Property and Land Taxation for Inclusive Local Finance}.
IMF Fiscal Affairs Department, Washington, DC.

\bibitem[Ingram and Hong(2023)]{IngramHong2023}
Ingram, G. K., \& Hong, Y.-H. (2023).
\textit{Value Capture and Land-Based Financing of Urban Infrastructure}.
Lincoln Institute of Land Policy.

\bibitem[Kopczuk(2022)]{Kopczuk2022}
Kopczuk, W. (2022).
Property taxation: Efficiency and incidence.
\textit{Annual Review of Economics}, 14, 249–274.

\bibitem[Mills(1967)]{Mills1967}
Mills, E. S. (1967).
An aggregative model of resource allocation in a metropolitan area.
\textit{American Economic Review}, 57(2), 197–210.

\bibitem[Mirrlees(1971)]{Mirrlees1971}
Mirrlees, J. A. (1971).
An exploration in the theory of optimum income taxation.
\textit{Review of Economic Studies}, 38(2), 175–208.

\bibitem[Muth(1969)]{Muth1969}
Muth, R. F. (1969).
\textit{Cities and Housing}.
University of Chicago Press.

\bibitem[Niepelt(2014)]{Niepelt2014}
Niepelt, D. (2014).
Wealth taxation, revenue, and welfare in a small open economy.
\textit{Journal of International Economics}, 94(2), 198–210.

\bibitem[Oates and Schwab(1997)]{OatesSchwab1997}
Oates, W. E., \& Schwab, R. M. (1997).
The impact of urban land taxation: The Pittsburgh experience.
\textit{National Tax Journal}, 50(1), 1–21.

\bibitem[OECD(2023)]{OECD2023}
OECD. (2023).
\textit{Housing Taxation in OECD Countries}.
OECD Publishing, Paris.

\bibitem[OECD(2024)]{OECD2024}
OECD. (2024).
\textit{Land, Location and Public Finance: Emerging Practices in Land Value Capture}.
OECD Publishing, Paris.

\bibitem[Peterson(2009)]{Peterson2009}
Peterson, G. E. (2009).
\textit{Unlocking Land Values to Finance Urban Infrastructure}.
World Bank.

\bibitem[Plassmann and Tideman(2022)]{PlassmannTideman2022}
Plassmann, F., \& Tideman, T. N. (2022).
A reappraisal of the Pennsylvania experience with two-rate taxation.
\textit{Land Economics}, 98(4), 625–648.

\bibitem[Rognlie(2015)]{Rognlie2015}
Rognlie, M. (2015).
Deciphering the fall and rise in the net capital share.
\textit{Brookings Papers on Economic Activity}, Spring, 1–69.

\bibitem[Samuelson(1954)]{Samuelson1954}
Samuelson, P. A. (1954).
The pure theory of public expenditure.
\textit{Review of Economics and Statistics}, 36(4), 387–389.

\bibitem[Stiglitz(2015)]{Stiglitz2015}
Stiglitz, J. E. (2015).
The theory of taxation and public economics revisited.
\textit{NBER Working Paper}, No. 21624.

\bibitem[Tiebout(1956)]{Tiebout1956}
Tiebout, C. M. (1956).
A pure theory of local expenditures.
\textit{Journal of Political Economy}, 64(5), 416–424.

\bibitem[Vickrey(1977)]{Vickrey1977}
Vickrey, W. (1977).
\textit{Agenda for Progressive Taxation}.
MIT Press.

\bibitem[Wheaton(1998)]{Wheaton1998}
Wheaton, W. C. (1998).
Land use and density in cities with congestion.
\textit{Journal of Urban Economics}, 43(2), 258–272.

\bibitem[Zodrow(2001)]{Zodrow2001}
Zodrow, G. R. (2001).
The property tax as a capital tax: A room with three views.
\textit{National Tax Journal}, 54(1), 139–156.

\bibitem[Zodrow and Mieszkowski(1986)]{ZodrowMieszkowski1986}
Zodrow, G. R., \& Mieszkowski, P. (1986).
Pigou, Tiebout, property taxation, and the underprovision of local public goods.
\textit{Journal of Urban Economics}, 19(3), 356–370.

\end{thebibliography}
\end{document}